\documentclass[12pt]{iopart}

\usepackage{graphicx} 
\usepackage[official]{eurosym}
\usepackage[colorinlistoftodos]{todonotes}
\usepackage{iopams}  
\expandafter\let\csname equation*\endcsname\relax
\expandafter\let\csname endequation*\endcsname\relax
\usepackage{amsmath}
\usepackage{hhline}
\usepackage{color,soul}
\usepackage{siunitx}

\begin{document}

\title[A laser-plasma platform for photon-photon physics]{A laser-plasma platform for photon-photon physics}

\newcommand{\IC}{The John Adams Institute for Accelerator Science, Imperial College London, London, SW7 2AZ, UK}

\newcommand{\UMICH}{Center for Ultrafast Optical Science, University of Michigan, Ann Arbor, MI 48109-2099, USA}

\newcommand{\DESY}{Deutsches Elektronen-Synchrotron DESY, Notkestr. 85, 22607 Hamburg, Germany}

\newcommand{\QUB}
{School of Mathematics and Physics, Queen's University of Belfast, BT7 1NN, Belfast, UK}

\newcommand{\HIJENA}
{Helmholtz Institut Jena, Fröbelstieg 3, 07743 Jena, Germany}

\newcommand{\JENA}
{Institut f\"{u}r Optik und Quantenelektronik, Friedrich-Schiller-Universit\"{a}t, 07743 Jena, Germany}

\newcommand{\CLF}{Central Laser Facility, STFC Rutherford Appleton Laboratory, Didcot OX11 0QX, UK}

\newcommand{\YORK}{Department of Physics, University of York, York, YO10 5DD, UK}

\newcommand{\OXFORD}{Department of Physics, University of Oxford, Oxford, OX1 3PU, UK}

\newcommand{\AWE}{AWE Aldermaston, Reading RG7 4PR, UK}

\newcommand{\CELIA}{Université de Bordeaux-CNRS-CEA, Centre Lasers Intenses et Applications (CELIA), UMR 5107, F-33405 Talence, France}

\newcommand{\LBNL}{Lawrence Berkeley National Laboratory, Berkeley, CA 94720, US}

\newcommand{\LANC}{Physics Department, Lancaster University, Lancaster LA1 4YB, UK}

\newcommand{\CERN}{CERN, Geneva, Switzerland}

\newcommand{\imp}{$^1$}
\newcommand{\hijena} {$^2$}
\newcommand{\jena}{$^3$}
\newcommand{\qub}{$^4$}
\newcommand{\clf}{$^5$}
\newcommand{\york}{$^6$}
\newcommand{\desy}{$^{7}$}
\newcommand{\cern}{$^8$}
\newcommand{\oxf}{$^9$}
\newcommand{\umich}{$^{10}$}
\newcommand{\awe}{$^{11}$}
\newcommand{\celia}{$^{12}$}
\newcommand{\lbnl}{$^{13}$}
\newcommand{\lanc}{$^{14}$}

\author{
B.~Kettle\imp,
D.~Hollatz\hijena$^,$\jena,
E.~Gerstmayr\imp,
G.M.~Samarin\qub,
A.~Alejo\qub,
S.~Astbury\clf,
C.~Baird\york,
S.~Bohlen\desy,
M.~Campbell\cern, 
C.~Colgan\imp,
D.~Dannheim\cern,
C.~Gregory\clf,
H.~Harsh\hijena$^,$\jena,
P.~Hatfield\oxf, 
J.~Hinojosa\umich,
Y.~Katzir\clf,
J.~Morton\awe, 
C.D.~Murphy\york,
A.~Nurnberg\cern, 
J.~Osterhoff\desy,
G.~P\'erez-Callejo\celia,
K.~P\~{o}der\desy, 
P.P.~Rajeev\clf,
C.~Roedel\hijena,
F.~Roeder\jena,
F.C.~Salgado\hijena$^,$\jena ,
G.~Sarri\qub,
A.~Seidel\hijena$^,$\jena,
S.~Spannagel\cern$^,$\desy, 
C.~Spindloe\clf,
S.~Steinke\lbnl,
M.J.V.~Streeter\qub$^,$\lanc,
A.G.R.~Thomas\umich$^,$\lanc,
C.~Underwood\york,
R.~Watt\imp,    
M.~Zepf \hijena$^,$\jena,
S.J.~Rose\imp
~and~
S.P.D.~Mangles\imp
}

\address{\imp \IC} 
\address{\hijena \HIJENA} 
\address{\jena \JENA} 
\address{\qub \QUB} 
\address{\clf \CLF} 
\address{\york \YORK} 
\address{\desy \DESY} 
\address{\cern \CERN} 
\address{\oxf \OXFORD} 
\address{\umich \UMICH} 
\address{\awe \AWE} 
\address{\celia \CELIA} 
\address{\lbnl \LBNL} 
\address{\lanc \LANC} 

\ead{b.kettle@imperial.ac.uk, stuart.mangles@imperial.ac.uk}
\vspace{10pt}

\begin{abstract}
We describe a laser-plasma platform for photon-photon collision experiments to measure fundamental quantum electrodynamic processes such as the linear Breit-Wheeler process with real photons. 
The platform has been developed using the Gemini laser facility at the Rutherford Appleton Laboratory. 
A laser wakefield accelerator and a bremsstrahlung convertor are used to generate a collimated beam of photons with energies of hundreds of MeV, that collide with keV x-ray photons generated by a laser heated plasma target.
To detect the pairs generated by the photon-photon collisions, a magnetic transport system has been developed which directs the pairs onto scintillation-based and hybrid silicon pixel single particle detectors. 
We present commissioning results from an experimental campaign using this laser-plasma platform for photon-photon physics, demonstrating successful generation of both photon sources, characterisation of the magnetic transport system and calibration of the single particle detectors, and discuss the feasibility of this platform for the observation of the Breit-Wheeler process. 
The design of the platform will also serve as the basis for the investigation of strong-field quantum electrodynamic processes such as the nonlinear Breit-Wheeler and the Trident process, or eventually, photon-photon scattering.
\end{abstract}

%
%
%
%
%

\section{Introduction}
\subsection{Background}

Gregory Breit and John A. Wheeler predicted in 1934 that when two photons collide, they can produce an electron and a positron in a process called ``inelastic photon-photon scattering'' \cite{Breit1934}. 
This Breit-Wheeler (BW) process $ \gamma  +  \gamma  \rightarrow e^- + e^+ $ has been studied in detail using the theoretical framework of quantum electrodynamics (QED)~\cite{greiner2008quantum}.
In the linear BW process, two photons must have sufficient energy to create an electron and a positron, i.e. the invariant mass of the collision must satisfy $ \sqrt{s} > 2m_{\rm e}c^2$, where $\sqrt{s}$ is the total energy measured in the centre of momentum frame.
For two photons of energy $E_1$ and $E_2$ interacting  at angle $\theta$, $s=[1-\cos(\theta)]E_1 E_2 $.
To date, the linear BW process has not been observed in the laboratory with real photons due to the lack of suitably bright sources of photons with sufficient energy.
There have been proposals for $\gamma-\gamma$ colliders using conventional particle accelerators~\cite{Asner2003, Takahashi2019}, but none have yet to be realised.

Indeed, in their original paper, Breit and Wheeler even stated: ``It is hopeless to try to observe the pair formation in the laboratory'' as in 1934 there was no conceivable way of generating such photon sources. 
Today however, bright electron beams with GeV energies can be produced over a few centimetres by laser wakefield acceleration (LWFA)~\cite{Kneip2009, Gonsalves2019}. 
These electron beams can be used to produce brilliant $\gamma$  ray beams by bremsstrahlung~\cite{Glinec2005,lemos2018bremsstrahlung, dopp2016bremsstrahlung, underwood2020development} and inverse Compton scattering \cite{Sarri2014,Yan2017,Cole2018}.
Colliding such brilliant $\gamma$  ray photons with a suitable x-ray photon field could be a viable route to study two-photon physics in the laboratory for the first time~\cite{Pike2014, golub2021linear}.

The BW process plays  a pivotal role in a range of astrophysical phenomena \cite{Ruffini2010}.
The linear BW process is the key process in Pair-Instability Supernovae (PIS) and the closely related pulsational pair-instability supernova, the fate of massive stars where the BW process softens the equation-of-state at the core, leading to collapse and ignition~\cite{Rakavy1967,Barkat1967}. 
PIS are of particular interest as they might be the only stars observable in the epoch of reionization when the first stars are formed, meaning that they may be the only view into star formation in the metal-poor early Universe possible with the James Webb Space Telescope~\cite{Pan2012,Regos2020}.  
The BW process also occurs in the coronae of accretion disks in Active Galactic Nuclei~\cite{Bonometto1971}, shaping particle spectra~\cite{Fabian2015}. 
Finally, the BW process occurs when $\gamma$  rays interact with the extragalactic background light (EBL), producing a cut-off of $\gamma$  rays at $\sim10^{13}$ eV~\cite{Gould1966}. 
There has been some suggestion of the possibility of non-standard model physics in the spectra of blazars, which some studies suggest have anomalously hard high-energy $\gamma$  ray spectra after the BW process is accounted for, implying less pair-production from QED than anticipated~\cite{Horns2012,Galanti2020}. 
This problem could well be solved with conventional physics (e.g. corrections to models of the EBL~\cite{Biteau2015}) as happened for the \textit{`Compactness Problem'}, a contradiction that arose from $\gamma$  ray burst light curve data which was eventually solved by including relativistic motion in the calculations~\cite{Piran2004}.
Nonetheless this illustrates the importance of confirming that QED physics behaves as expected in this regime for this important process.

The existence of a photon-photon experimental platform studying the BW effect will serve as a basis for the investigation of other QED processes including elastic photon-photon scattering $\gamma\gamma\rightarrow\gamma\gamma$ and the nonlinear BW process.
The nonlinear BW process is closely related to the BW process but involves the interaction of one high-energy photon with many $(n)$ low energy ones, i.e. $ \gamma + n \omega  \rightarrow e^- + e^+$. 
The generation of matter via the nonlinear BW process plays a large role at the poles of rotating black holes \cite{Blandford1977}.
In 1997, an experiment at SLAC collided a 48 ~GeV electron beam with a then state-of-the art laser pulse \cite{Bula1996,Burke1997}, demonstrating the nonlinear BW process, thus studying the perturbative regime of strong-field quantum electrodynamics~\cite{Bamber2004}.

Elastic photon-photon scattering, $\gamma\gamma\rightarrow\gamma\gamma$, is important in models of primordial abundances and may also affect the observed spectra from distant gamma-ray bursts ~\cite{Ellis1992, Svensson1990}. 
Photon-photon scattering involving virtual photons has previously been observed. 
Delbr{\"u}ck scattering, $\gamma\gamma*\rightarrow\gamma\gamma*$, where a real photon scatters from a virtual photon in the electric field of an ion, was observed in 1975~\cite{Schumacher1975} and scattering between two virtual photons in the Coulomb field of two high energy ion beams $\gamma*\gamma*\rightarrow\gamma\gamma$ was observed at the Large Hadron Collider in 2017~\cite{ATLAS2017}.
However, photon-photon scattering is most relevant in astrophysics with real photons close to threshold for pair production energy ($\sqrt{s} \leqslant 2m_e c^2$) as at higher energies pair production dominates. 
Photon-photon scattering with real photons has yet to be observed in the laboratory.

\subsection{Laser-plasma approach}
In 2014, Pike et al.~\cite{Pike2014} proposed an ambitious scheme that could enable the observation of the linear BW process by combining a laser-driven $\gamma$ beam source with a laser-driven x-ray source. 
They proposed using a megajoule laser pulse, e.g. the National Ignition Facility~\cite{Miller2004NIF}, coupled to a petawatt short pulse laser to produce 10,000 pairs in a single shot. 
However, no such combination of high-power lasers exists worldwide.
Ribeyre et al. (2016) \cite{Ribeyre2016} proposed the interaction between two high-power laser generated $\gamma$-ray beams.
However this method relies on extremely precise spatial and temporal overlap of the two beams, and due to the approximately equal incoming photon energies, would generate pair particles into a wide range of angles, making detection difficult.
In this article we present an alternative laser-based approach that can operate at existing laser facilities, benefits from directed pair production, and which can make use of higher shot repetition rates to provide the statistics needed for any given measurement.

Our approach is based on the interaction of a single $\gamma$  beam colliding with an x-ray bath, similar in nature to that of Pike et al.~\cite{Pike2014}, but with a few hundred terawatt laser system providing tens of joules (as opposed to a megajoule) at a repetition rate of over 100 shots an hour (as opposed to a few shots a day).
A major difficulty that a laser based approach overcomes is the spatial and temporal overlap that goes hand-in-hand with colliding intense photon sources.
Using the same laser system to drive the photon sources means that they can be inherently synchronised in time (within tens of femtoseconds) and do not suffer from significant temporal jitter.
Making use of a large x-ray radiation field ($>$\,\SI{100}{\micro\metre}\,$\times$\,\SI{100}{\micro\metre} and tens of picoseconds in duration) means that the spatial alignment of the two photon sources is also relatively straightforward.
In addition to the overlap, the strong asymmetry of the photon energy of the two sources also results in a strongly directional emission of the created electron-positron pairs. 
This is particularly important considering that for many  experimental configurations the BW process can be obscured by other more efficient pair creation processes, such as the Bethe-Heitler process,  when high energy photons interact with the matter in the vicinity of the photon-photon collision. 
A strategic and tailored detection system that can decouple the BW signal from background pairs from other processes is required, and a clean particle-free interaction volume is desirable.
Two decoupled photon sources with a spatially directed signal is a  solution to this problem and presents a major advantage over other suggested schemes that collide two sources of photons that are equal in energy~\cite{Drebot2017PRAB}.

Although the predicted pair creation numbers per shot are lower than that of Pike et al. (around unity, as discussed later), many shots can be taken per day to build a statistical significance.
It is also worth noting that the platform we present here can also be modified to study the nonlinear BW process by colliding the $\gamma$ ray photons directly with the intense optical laser field that would otherwise drive the x-ray bath.

In this paper we detail the design and commissioning of a laser plasma platform for photon-photon physics. 
Our characterisation and modelling of the background levels and expected signal level  indicate that a measurement of the linear BW process is feasible within a reasonable timeframe (1000s of laser shots), but show that this is very sensitive to the performance of the photon sources.  
Further developments of the platform are discussed, which should bring the number of required shots to less than a few hundred.
\section{Experimental platform overview}

\begin{figure} 
   \centering
   \includegraphics[width=0.95\textwidth]{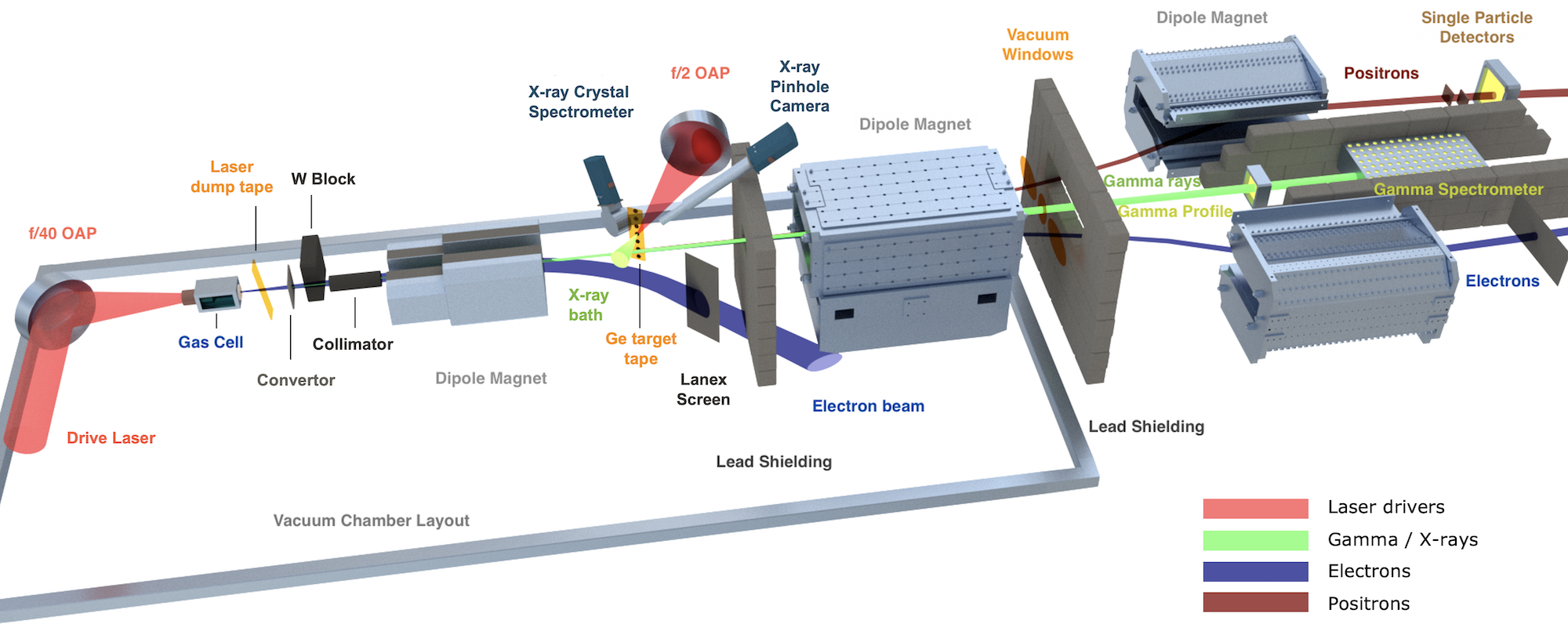} 
   \caption{The experiment setup. An f/40 laser pulse was used to drive a laser-wakefield accelerated electron beam in a gas cell. This high-energy electron beam was used to create on axis $\gamma$-rays before being swept away by a magnet. A second laser pulse was incident on a germanium target tape, driving an x-ray bath through which the $\gamma$ beam passed. Any particle pairs created in this interaction region propagate on axis and are swept away to single particle detectors using a magnetic transport system.}
   \label{fig:ExpSetup}
\end{figure}

The platform was developed at the Gemini laser system for strong field QED experiments \cite{Keitel_Gemini2010} at the Central Laser Facility, UK and adapted to the specific requirements of this experiment.
It is an optically synchronised two beam system, where each beam contains up to 15 J in $\approx45$ fs (FWHM), with a shot rate of once every 20 seconds.
The synchronisation of the two beams allows for a temporal overlap and jitter of less than 10~fs~\cite{corvan2016optical}.
One of the beams was used in a long focal length $f$/40 configuration to drive a laser-wakefield accelerator.
The high energy electron bream produced in this interaction is subsequently used to generate  bremsstrahlung MeV $\gamma$ photons.
The second beam was stretched in duration to 40 ps (FWHM), and used to drive the laser-plasma generated keV x-ray field (using an $f$/2 focusing optic).
These two sources are made to overlap spatially within the collision interaction point and a magnetic transport system relays any created electron-positron pairs to a set of single particle detectors.
An overview of the experimental layout can be seen in figure~\ref{fig:ExpSetup} and a full description of the setup is as follows~\cite{DataRepo}.

The long focal length arm of the laser, which drives the laser-wakefield accelerator, is focused onto a gas cell.
In this case the cell was 17.5 mm long and filled with helium and 2\% nitrogen as a dopant at an electron density of $(2.6\pm 0.4) \times 10^{18}\,\mathrm{cm}^{-3}$.
The typical measured laser spot size was \SI{44}{\micro\metre}\,$\times$\,\SI{53}{\micro\metre} FWHM at a pulse duration of $45 \pm 5$ fs and $5.5 \pm 0.6$ J on target, which corresponds to a normalised vector potential $a_0 = 1.1 \pm 0.2$.
As the laser pulse travels through the gas it strips electrons from the ions, forming a plasma.
The ponderomotive force of the laser pulse pushes these electrons away from the intense pulse, while the ions remain stationary, forming a positively charged cavity that trails in the wake of the laser pulse~\cite{TajimaDawson1979}.
The electrons can then be trapped and accelerated in the cavity to relativistic energies, travelling in the direction of the laser pulse. 
After driving the laser-wakefield accelerator, the remaining drive laser exits the gas cell and is removed from the beamline using a refreshable tape drive.

\begin{figure} 
   \centering
   \includegraphics[width=0.95\textwidth]{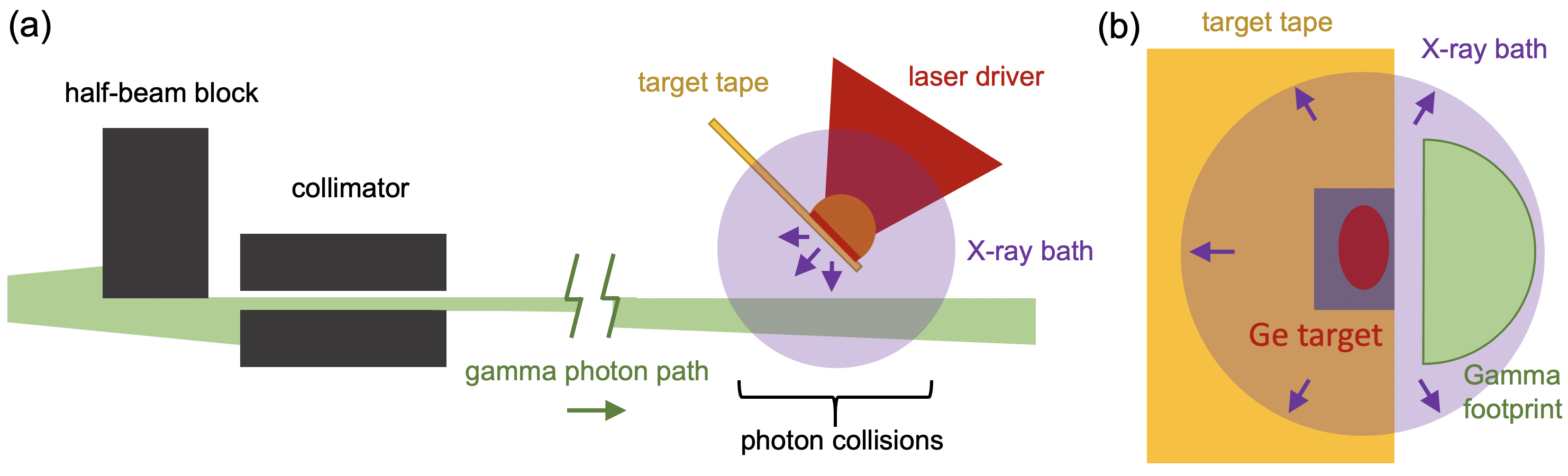} 
   \caption{(a) Simplified view of the photon collision geometry. 
    The tungsten half-beam block and collimator shape the $\gamma$ beam to the desired profile before, much further downstream, colliding with the pre-expanded x-ray bath. 
    (b) On-axis view of the overlap geometry. 
    The x-ray bath radiates in all directions from the germanium target dot, providing a suitable particle-free collision volume for the D-shaped $\gamma$ beam to pass through.}
   \label{fig:OverlapGeometry}
\end{figure}

The generated high-energy electron beam continues along the beamline where it passes through a 1 mm thick bismuth \textit{`converter'} target.
While passing through the high-Z atoms of the converter, the electrons generate bremsstrahlung radiation, emitting $\gamma$ photons with over 100 MeV energy along the propagation direction.
The $\gamma$ beam is collimated to $\sim$ 8 mrads divergence using a 10 cm long tungsten cylinder with a 2 mm aperture.
One half of the $\gamma$ beam is also blocked by a 5 cm thick tungsten shield.
The remaining footprint of the $\gamma$ beam is semi-circular in shape.
This D shape is designed to allow the tape target of the x-ray bath source (further downstream) to be placed in close proximity to the centre of the $\gamma$ beam, but hidden from direct photon collisions at the interaction point.
See figure~\ref{fig:OverlapGeometry} for a depiction of the overlap geometry.

After the $\gamma$ beam has been tailored to the desired shape, a $\int B dx = 0.4~\mathrm{Tm}$ dipole magnet is used to sweep away the initial electron beam as well as any secondary particles generated from collisions with the tungsten shield elements.
This combination of $\gamma$ beam tailoring and particle removal provides a clean environment for photon-photon collisions.

The second laser beam is incident upon a plastic tape containing regularly spaced germanium targets, which are placed 1~mm from the $\gamma$ beam footprint.
The laser pulse rapidly heats and ionises the metal, which emits M-L shell transition radiation in the 1 - 2 keV region; see section~\ref{section:xrays} for more details.
This emission radiates in all directions, but we are concerned with the x-ray plume that penetrates through the plastic and out of the rear of the target towards the interaction region.
The two drive laser pulses should be timed so that the $\gamma$ beam passes through the x-ray field at its highest photon density.
The x-ray driver is also short enough in duration that no ions from the germanium plasma plume have penetrated through to the collision volume by the time the $\gamma$ beam is incident.
The \SI{5}{\micro\metre} thick Kapton tape substrate helps prevent plasma from the front (laser-irradiated) side of the germanium target penetrating through to the rear (where photon collisions occur).
For ions to propagate from the plasma to the interaction volume (1 mm away) within 100 ps of the laser driver, they would need to move at $\sim10^7$ m/s, or $\sim3\%$ the speed of light.
However calculations of the laser-plasma interaction indicate a peak electron temperature of $\sim2.5$ keV, near the peak of the pulse, with an associated ion sound speed ($c_s\sim3.5 - 6.3\times10^5$ m/s) that is a fraction of that required to reach the interaction by the collision time.
Additional hydrodynamic simulations support this, showing that the Kapton substrate of the targets has not been burnt through by the time the $\gamma$ beam collides with the interaction volume.
Because the $\gamma$-rays are much higher energy than the x-rays, any electrons and  positrons produced in the photon-photon collisions travel along the initial $\gamma$-ray beam direction. 
A second dipole magnet $\int B dx = 0.35~\mathrm{Tm}$  with a large aperture sweeps the generated positrons off axis.
It is at this point the positrons exit the vacuum chamber through a \SI{125}{\micro\metre} Kapton window.
They pass through  a third dipole magnet with opposite polarity to the second, redirecting the particles along a focused beam path.
This particle transport system is discussed in detail in section~\ref{section:transport}.

Single particle detectors (SPDs) are placed on this beam path to measure the particle flux.
Electrons generated within the interaction region are swept by the second dipole in the  opposite direction, and a fourth magnet (identical to the third one) directs the electrons to a beam dump. 
The profile and spectrum of the on-axis $\gamma$-ray beam are diagnosed using scintillator-based detectors at the end of the beam line. 
The $\gamma$-ray diagnostics and single particle detectors are discussed in more detail in section \ref{section:gammas} and section \ref{section:detectors} respectively.
Lead shielding is used extensively throughout the setup to mitigate the influence of unwanted high energy particle noise on the diagnostics, details of the shielding and the effect on background levels are discussed in section \ref{section:predictions}.  
\section{$\gamma$-ray photon source}\label{section:gammas}

\begin{figure}
    \centering
    \includegraphics[width=0.95\textwidth]{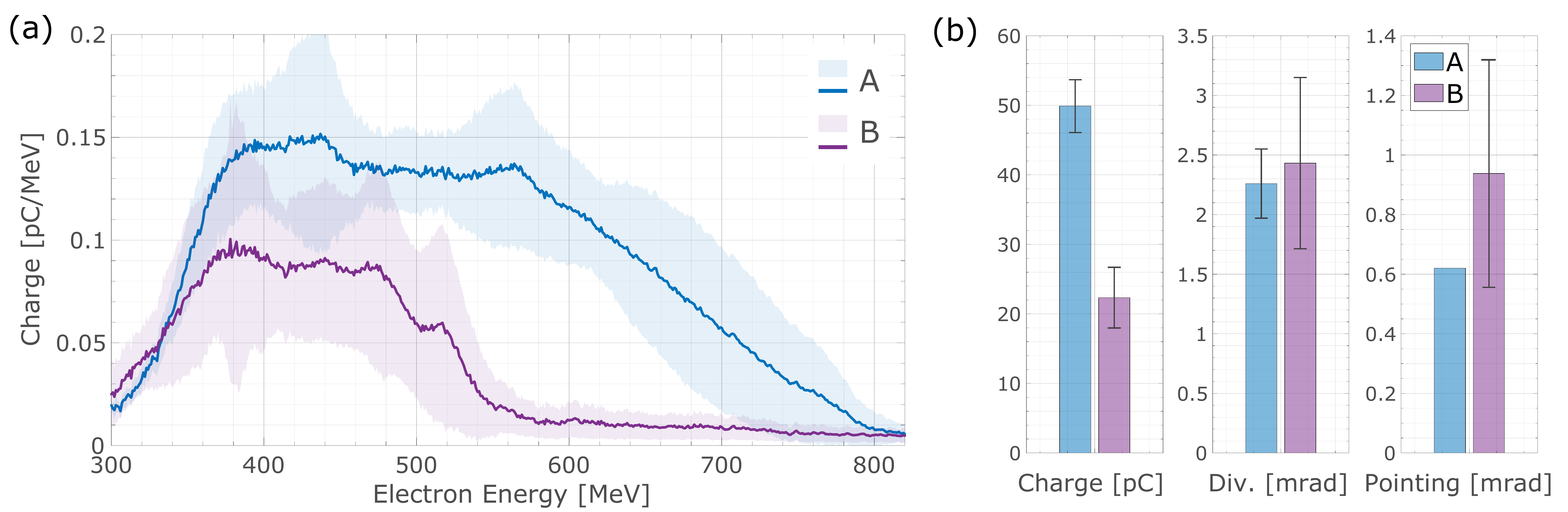}
    \caption{Properties of the electron beam used to generate the $\gamma$-ray photon source: (a) Average electron spectra with standard deviation from mean (shaded). 
    (b) Average electron beam properties with standard deviations (error bars). 
    A (blue) and B (purple) represent the optimum beam performance and performance during commissioning, respectively. }
    \label{fig:ElectronPerformance}
\end{figure}

During the commissioning of the platform the laser energy used to drive the accelerator was limited to 5 J, which is significantly below the maximum that can be achieved in Gemini (12-15 J on target).  
With 5 J laser energy the best electron beam performance achieved (data set `A') reached on average a maximum energy of $710\pm50$ MeV and contained a charge of $50\pm 7$ pC.
The data set contained 10 shots and the average spectra of these beams is presented in figure~\ref{fig:ElectronPerformance} (a), labelled dataset A (blue).
The measurement had a low-energy cut off at $\sim350$ MeV due to the yoke of the spectrometer magnet.
The beams had a divergence of $2.3\pm0.3$ mrad and a shot-to-shot pointing standard deviation of $0.6$ mrad.
Additionally, during the commissioning phase of the platform (discussed later), the LWFA electron beam performed in a reduced capacity, with lower maximum energies achieved and less than half the total charge (data set `B').
There were 79 shots in total and the average spectra for these runs are provided in figure~\ref{fig:ElectronPerformance}, labelled dataset B (purple).

\begin{figure}
    \centering
    \includegraphics[width=0.95\textwidth]{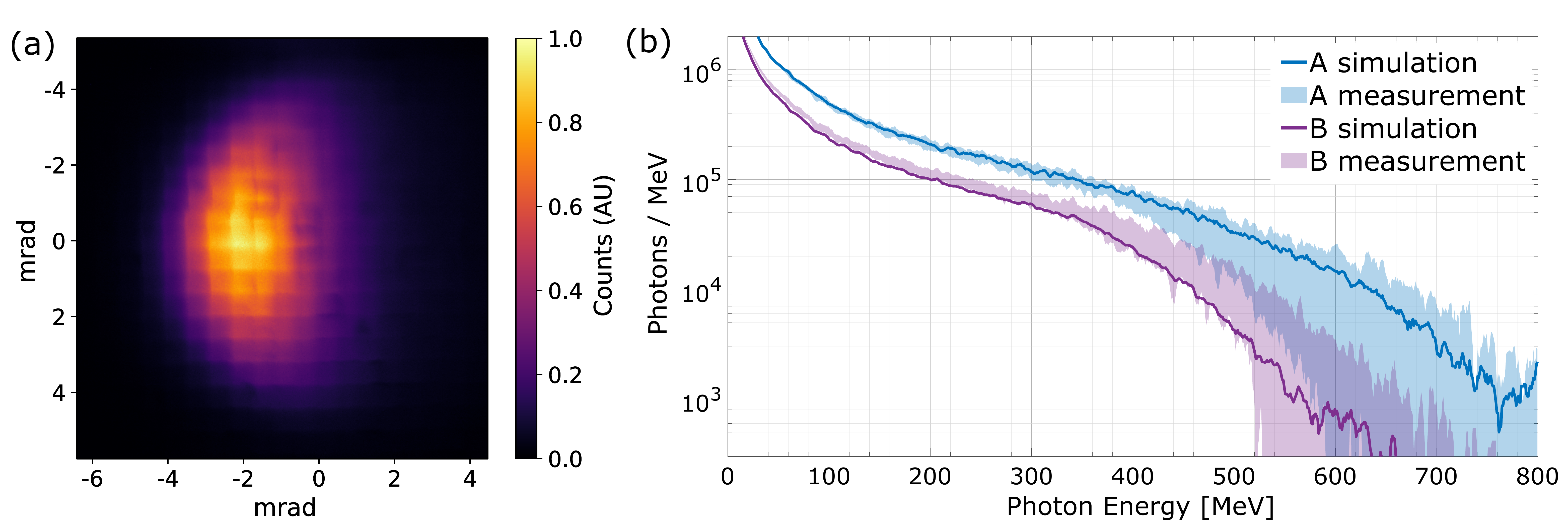}
    \caption{Properties of the $\gamma$-ray photon source: (a) $\gamma$ beam profile example. The D-shaped footprint produced by the collimator and half-block shadows is clear. 
    (b) Simulated $\gamma$ spectra for both data sets (solid lines) and $1\sigma$ interval of the first 10 spectra of the corresponding data sets measured with the $\gamma$ spectrometer (shaded regions).}
    \label{fig:GammaPerformance}
\end{figure}

The $\gamma$ profile diagnostic took the form of an array of $20 \times 20$ caesium-iodide (CsI) crystals, each of dimensions $2\,\mathrm{mm} \times 2\,\mathrm{mm}$ and $20\,\mathrm{mm}$ thick, positioned normal to the $\gamma$ beam and $3.39\,\mathrm{m}$ downstream of the $\gamma$ source. 
The scintillation of the CsI crystals upon exposure to high energy photons is observed with an optical camera.
An example $\gamma$ beam profile from this diagnostic can be seen in figure~\ref{fig:GammaPerformance} (a), the D shaped profile due to the combination of the collimator and tungsten block shielding the x-ray target are clearly apparent.
Another larger array of $33 \times 47$ crystals, each $5\,\mathrm{mm} \times 5\,\mathrm{mm} \times 50\,\mathrm{mm}$, is positioned along the beam axis behind the profile diagnostic and measures the decay of signal in the propagation direction to deduce the spectrum \cite{Behm2018}.
The $\gamma$ spectrum is inferred using a forward model to calculate the energy deposited in each crystal, finding the model parameters which most closely reproduce the observed signal using a least squares approach.
In this case our forward model is based on Geant4 simulations of the signal produced due to the $\gamma$ rays produced by a wide range of experimentally observed electron spectra.
Both the profile and spectrum scintillator arrays are observed by cooled 14-bit EMCCD cameras (Andor iXon), with single photon sensitivity to the scintillation light.

The converter target scatters the incident electron beam such that the properties of the $\gamma$ beam and the electrons cannot be measured simultaneously. 
Therefore the electron beam was characterised without the converter in place, and this was subsequently used to simulate the expected average $\gamma$ spectrum.
The $\gamma$ spectra that were measured with the larger crystal array immediately after the characterisation of the electron beam (figure~\ref{fig:GammaPerformance} (b), shaded regions) agreed well with the simulated spectra (figure~\ref{fig:GammaPerformance} (b), solid lines).

Geant4 simulations using dataset A indicate the emission of $(3.5 \pm 0.5) \times 10^{8}$ photons or $(7 \pm 1) \times 10^6$ photons/pC, distributed in an exponentially falling spectrum with a high-energy cutoff at the maximum electron energy (see figure~\ref{fig:GammaPerformance} (b)).
$(7 \pm 3)\times 10^{7}$ of these photons, or about $20\%$ of the total photon yield, have an energy above $100$~MeV.
By matching the average electron beam and the average yield on the $\gamma$ profiler, we obtain a calibration factor of $(5.0 \pm 0.4) \times 10^6$ $\gamma$-counts/pC. 
This calibration allows us to estimate the electron beam charge on shots where the convertor target is in place (and hence no electron spectrometer measurement is possible).
The variations in the $\gamma$ yield indicate stronger fluctuations in the photon numbers from shot to shot than expected from the electron charge. 
This is believed to be due to pointing fluctuations and the limited size of the $\gamma$ detector occasionally missing outer portions of the beam.
As described previously, for photon-photon collisions, the $\gamma$ beam is collimated and one half is blocked to prevent interactions with the x-ray target (see figure~\ref{fig:OverlapGeometry} and \ref{fig:GammaPerformance} (a)).
This reduces the footprint of the beam and the effective yield of photons by a factor $3.6 \pm 1.2$. 

\begin{figure}[ht!]
    \centering
    \includegraphics[width=0.95\textwidth]{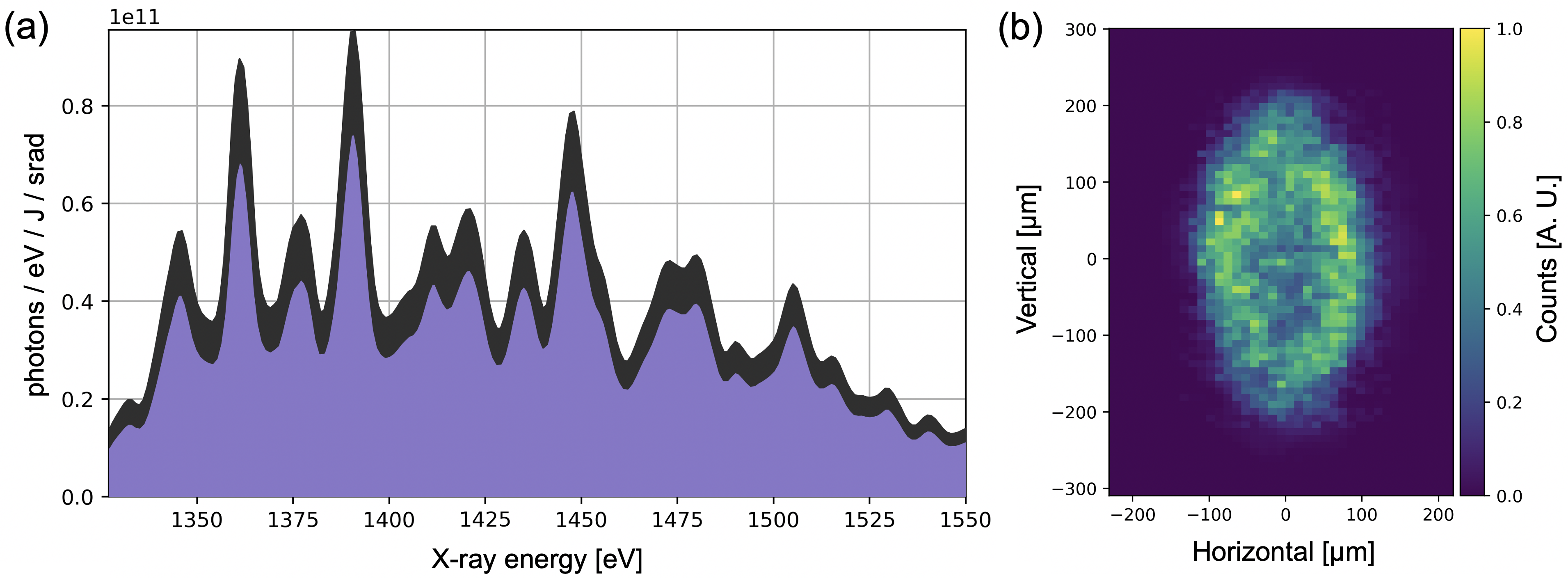}
    \caption{ (a) Average x-ray spectra emitted by the germanium target. Measured using a TlAP crystal spectrometer over 47 shots. The standard deviation is given by the black shaded area.
        (b) Example median-filtered x-ray pinhole camera image of the germanium target. The aspect ratio of the x-ray source closely resembles the $f/2$ focal spot.
        }
    \label{fig:XraysOverview}
\end{figure}


\section{x-ray photon source}\label{section:xrays}

The second arm of the laser system was used to generate the dense keV x-ray bath at the interaction point. 
The drive laser was focused onto a refreshable Kapton tape target system using an $f/2$ off-axis parabola (OAP).
Prior to reflecting from the final focussing optic, the 150 mm diameter beam passed through a distributive phase plate~\cite{Kato1984}.
This allowed the laser energy to be deposited in a large quasi-flat-top focal spot \SI{210}{\micro\metre}\,$\times$\,\SI{90}{\micro\metre} (major and minor diameters at $1/e^{2}$). 
Approximately $72\%$ of the laser energy is contained in this spot, with the remaining light being diffracted away into higher orders.
The laser pulse was also stretched to $40$ ps duration, resulting in an intensity of $2 \times 10^{15}$ Wcm$^{-2}$ on target.
The targets themselves were $\sim100$ nm coating of germanium metal on a 5\,$\pm$\,\SI{2}{\micro\metre} etched square window ($\sim1~\textrm{mm}^2$) of the Kapton tape (which is otherwise nominally \SI{25}{\micro\metre} thick)~\cite{SciTechTape}.
As the laser pulse is incident on the germanium coating, a hot dense plasma is created on the front surface.
In this highly-ionised plasma, M-L shell transitions in the germanium efficiently produce photons in the range of $1-2$ keV~\cite{PhillionHailey1986}.
These x-rays are emitted into a 4$\pi$ sphere, such that a few milimetres from the target a dense x-ray bath exists.
A similar technique has previously been used to rapidly heat targets to extreme conditions~\cite{Kettle2015}.
The Kapton target substrate thickness was optimised to increase the transmission of these keV photons, while suppressing sub-keV photons and preventing any material contamination to the interaction region.

A Thallium Acid Phthalate (TlAP) crystal spectrometer viewed the front of the tape targets at angle of $\approx55^\circ$ (to normal) to diagnose the x-ray yields.
Bragg reflections from the flat TlAP crystal onto an in-vacuum x-ray CCD (Andor DX-420-BN) allowed the x-ray spectrum to be measured between $1.3-1.5$ keV with a spectral resolution of 4 eV.
The average measured spectra from 47 shots can be seen in figure \ref{fig:XraysOverview} (a).
The absorption edge from an aluminium filter at $1.56$ keV provided a spectral marker allowing the dispersion across the CCD to be calculated.
A variety of other filters were also used, allowing the signal to be corrected for background contributions from direct CCD irradiation, higher order Bragg reflections and crystal-substrate fluorescence.

In addition to the crystal spectrometer, a 2-channel pinhole imaging system was employed. 
By imaging the x-ray producing plasma, the source size and target alignment were monitored on-shot.
An example image from this diagnostic can be seen in figure \ref{fig:XraysOverview} (b).
While the entire emission collected by the pinhole camera passed through a \SI{25}{\micro\metre} Be filter (acting as a $600$ eV high-pass filter), both channels of the imaging system had different spectral filters.
This allowed for a broad x-ray yield quantification, which correlated well with the total energy measured by the crystal spectrometer.

It was found that the total conversion efficiency from laser energy to $1.3 - 1.5$ keV x-rays was $(2.4 \pm 0.3) \%$. 
This corresponds to $(3.7 \pm 0.4)\times 10^{10}$ photons/eV/J/srad emitted normal to the front surface of the germanium target.
Taking into account the absorption in the kapton layer on the rear side of the targets, at the interaction region (1 mm from the tape) this corresponds to a photon density of ${(1.4 \pm 0.5) \times 10^{12}~\textrm{mm}^{-3}}$. 
These photon numbers assume the Kapton tape remains cold throughout the interaction, and are supported by simulations of the X-ray source directionality and duration ($97 \pm 11~\textrm{ps}$, FWHM). 
The average transmission of the Kapton tape is 47\% over the range $1.3 - 1.5~\textrm{keV}$.
The flux stability from shot-to-shot was found to have a standard deviation of  $28\%$.
A tape drive was used to place a fresh target at the focal plane of the x-ray drive laser. 
The tape targets were able to be aligned at a repetition rate of 1 shot per minute.
Note however that it should be possible to increase this rate with improvements to the target alignment system.
\section{Positron Transport System}\label{section:transport}

\begin{figure}
    \centering
    \includegraphics[width=0.95\textwidth]{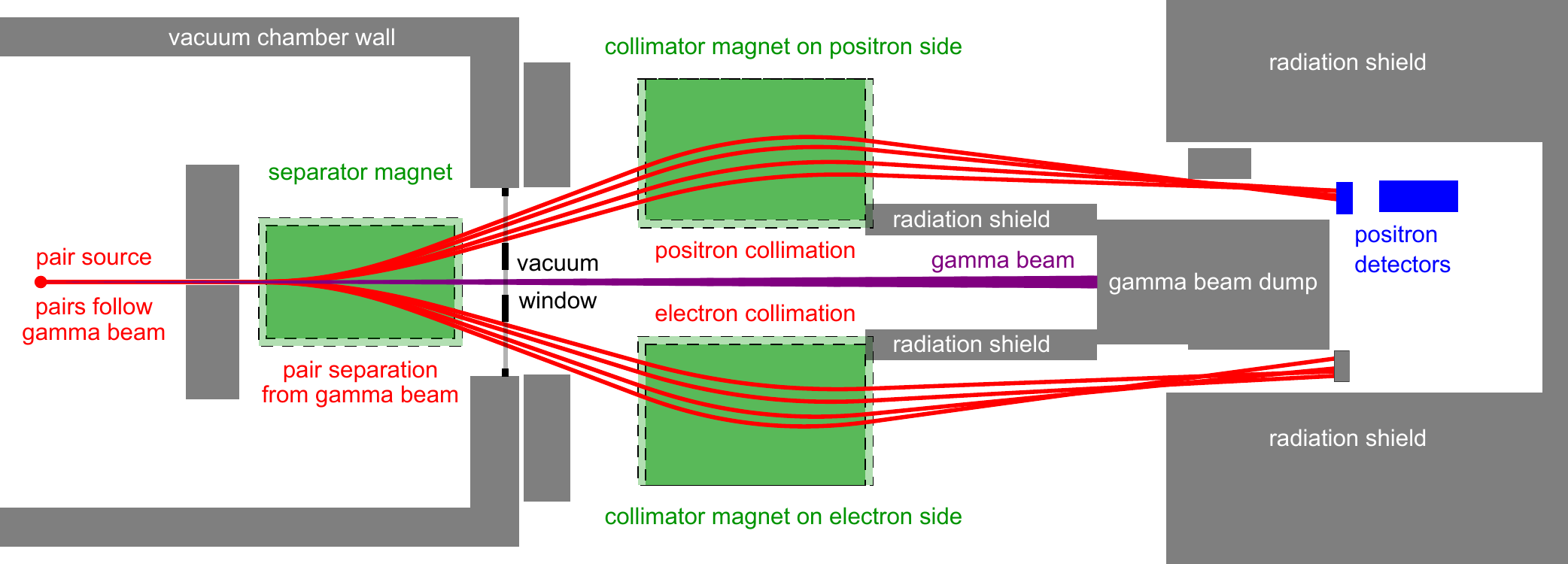}
    \caption{Overview of the analyser magnet system (AMS) to transport generated pairs to the single particle detectors (SPD). The combination of separator and collimator magnets was adapted for the energy range in this experiment by adjusting the length of the magnets and their separation. The electron and positron paths were adjusted to avoid the SPDs being directly exposed to radiation from noise sources inside the vacuum chamber (such as the electron source and convertor target).}
    \label{fig:TransportOverview}
\end{figure}

\begin{figure}
    \centering
    \includegraphics[width=0.95\textwidth]{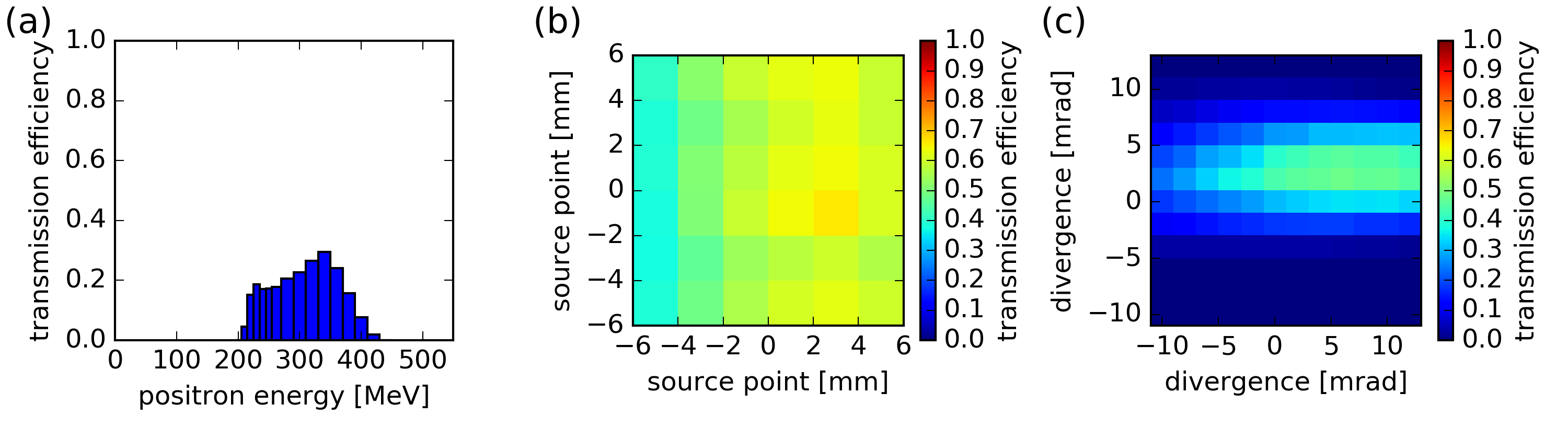}
    \caption{Transmission efficiency of positrons through the AMS to the CsI SPD depending on positron energy (a), source point (b) and divergence (c). For BW positrons, the integrated efficiency is 38\% above 200~MeV and 14\% overall.}
    \label{fig:TransportEfficiency}
\end{figure}

In order to measure the produced pairs, a system composed of permanent magnets and single particle sensitive detectors has been developed \cite{Keitel_Gemini2010} and specifically adapted for the energy range in this experiment.
The overall geometry is shown in figure~\ref{fig:TransportOverview}.
The generated pairs initially follow the $\gamma$-ray beam path, but after passing through an aperture in a lead wall, they were separated from the beam using a dipole magnet.
This magnet was 60\,cm long with a field strength of 0.6\,T ($\int B dx = 0.36~\mathrm{Tm}$) and is referred to as the separator magnet.
To avoid scattering of $\gamma$-rays on the poles, the magnet had a gap of 10\,cm.
The pairs were dispersed and exited the vacuum chamber through a \SI{125}{\micro\metre} thick Kapton window.
The pairs were then detected by two different types of single particle detectors, which have a limited spatial extent and are presented in section~\ref{section:detectors}.
To increase the collection efficiency of the generated pairs, two additional dipole magnets were used to focus the electrons and positrons to individual detectors.
These magnets are referred to as the collimator magnets.
They had an aperture of 9\,cm, were 70\,cm long and provided a magnetic field strength of 0.5\,T ($\int B dx = 0.35~\mathrm{Tm}$).
The combination of separator and collimator magnets is called the analyser magnet system (AMS).
The efficiency of this transport system is limited by various apertures along the beam path, especially the vacuum exit window frame which has three openings, one for each particle beam and a central one for $\gamma$-rays, see figures~\ref{fig:ExpSetup} and \ref{fig:TransportOverview}.
The AMS transport efficiency was evaluated using simulations performed with Radia \cite{elleaume1997computing}\cite{chubar1998three} for Wolfram Mathematica.
The simulations model the magnetic fields and calculate the trajectory of positrons through the system for a range of possible source positions and initial momenta. 
The collection of all trajectories represents a 5-dimensional function (the source point is fixed along the beam axis) that maps any positron to the detector plane depending on its initial phase space coordinate at the time of pair creation.
Combined with a check if individual trajectories are blocked by apertures, this function, which is called the Radia filter, determines if and where a positron hits the detector.
This is used to evaluate the transmission efficiency and the spatial SPD signal distribution for any incident positron population.
This filter was used to map positrons produced by a 1\,mm thick plastic PTFE calibration target to the SPD plane as a detector calibration, which is described in greater detail in section~\ref{section:detectors}.
Since it is a 5-dimensional filter, the transport efficiency is hard to visualise due to correlations between all 5 initial coordinates and also depends on the phase space distribution of the incident positrons.
A projection of the full filter to the energy axis as well as source point and divergence planes is shown on figure~\ref{fig:TransportEfficiency}, which exhibits a narrow spectral acceptance between 200 and 420\,MeV as well as a narrow angular acceptance between -1 and +7~mrad relative to $\gamma$ beam axis.
The source point does not affect the transmission significantly over the considered range from -6 to 6~mm relative to $\gamma$ beam axis.
The overall transport efficiency for positrons from the PTFE calibration target above 200\,MeV is 28\,\% and the transport efficiency for produced BW positrons above 200\,MeV is 38\,\%.
For all BWpositrons in the full spectral range, the transmission is 14\,\%.
The accuracy of calculated positron impact locations is approximately 1~cm at the detector plane, limited by the positioning accuracy of magnets, detectors and apertures inside the laboratory. 

\section{Single Particle Detectors}\label{section:detectors}

\subsection{Caesium Iodide Single Particle Detector}

A scintillation based single particle detector (SPD) system was employed.
It was composed of a thallium doped caesium iodide (CsI) crystal array coupled to a 4Picos ICCD from Stanford Computer Optics.
The crystals were coated with titanium dioxide to avoid optical crosstalk between neighbouring crystals.
The surface facing towards the camera was polished.
The 4Picos camera achieve single photon sensitivity through a multi-channel-plate with photoelectron multiplication factors up to $10^6$.
This combination of CsI crystals and 4Picos camera gives the SPD single particle sensitivity.

\begin{figure}
    \centering
    \includegraphics[width=0.95\textwidth]{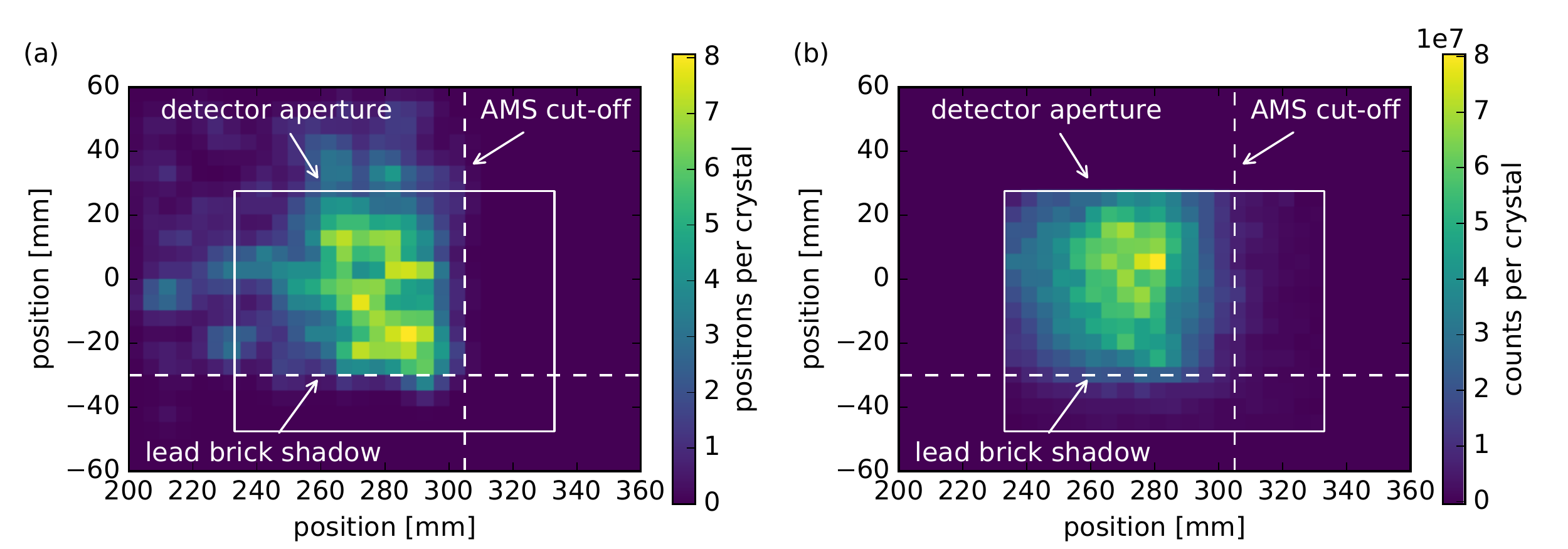}
    \caption{(a) The simulated positron hit map on the detector plane produced by the PTFE calibration target on a 9.2~pC$_r$ shot. (b) The measured detector response with the PTFE calibration target in the beam. The combined information from both images is used as a detector calibration.}
    \label{fig:simulated-vs-measured-PTFE-positron-distribution-on-SPD}
\end{figure}

The CsI SPD was calibrated with a PTFE calibration target in the beam close to the BW pair source point to produce a known positron signal on the SPD.
As the $\gamma$ beam passes through the PTFE, electron-positron pairs are generated in a narrow on-axis cone via the Bethe-Heitler process.
The PTFE positron population was calculated with Geant4 using the measured LWFA beam as input and simulating the bremsstrahlung process, $\gamma$-ray beam shaping and Bethe-Heitler pair production.
The measured electron beam performance (and subsequent Geant4 input) for these calibration shots are given in figure~\ref{fig:ElectronPerformance} dataset B.
The obtained positron population was filtered and projected to the detector plane using the AMS phase space filter calculated with Radia as described in section~\ref{section:transport}.
The simulated spatial distribution of positron impact locations on the SPD plane is shown on figure~\ref{fig:simulated-vs-measured-PTFE-positron-distribution-on-SPD} together with the measured signal.
The square box indicates the detector aperture and the dashed lines show positron cut-off's, one due to lead shielding that intentionally shadows the lower portion of the SPD and another one due to the AMS field distribution.
The regions where no positrons impact the SPD are used to measure background radiation other than positrons transported through the AMS, which is described in section~\ref{section:commissioning}.

The simulated positron yield integrated over the detector aperture is $(200\pm7)$ positrons$_{ams}$, where the error is the standard error of the mean over 3 identical Geant4 simulations.
The subscript $ams$ indicates that we are referring to positrons transmitted through the AMS with the associated properties specified in figure~\ref{fig:TransportEfficiency}.
The LWFA beam spectral shape is assumed to be constant for all calibration shots, which allows us to assume a linear dependency between positron yield and LWFA charge.
This normalisation to LWFA beam charge assuming the reference spectral shape is denoted with pC$_r$.
The Geant4 simulations were performed with a total charge of 9.2\,pC$_r$, so the positron yield produced by the PTFE target is obtained as $(21.7 \pm 0.8)~ \mathrm{positrons_{ams}}/\mathrm{pC_r}$.

The mean measured signal integrated over the detector area is $(8.44 \pm 0.69)\times 10^8$~counts$_n$, where the subscript $n$ denotes counts that are corrected for camera and imaging system effects.
The PTFE data was taken with an average charge of $(19.3 \pm 1.1)~\mathrm{pC_r}$, so the total measured PTFE positron yield is obtained as $(4.77 \pm 0.17)\times 10^7~\mathrm{counts}_n/\mathrm{pC_r}$ with errors being standard errors of the mean.
This is then corrected for background radiation by taking shots without the PTFE target in the beam, which was measured to be $(0.109 \pm 0.004)\times 10^7~\mathrm{counts_n}/\mathrm{pC_r}$, and gives the BG corrected yield of $(4.66 \pm 0.17)\times 10^7~\mathrm{counts_n}/\mathrm{pC_r}$.
This measured signal is now directly correlated to the simulated number of incident positrons to obtain the SPD calibration of $(2.15 \pm 0.03)\times 10^6~\mathrm{counts_n}/\mathrm{positron_{ams}}$.

\subsection{Timepix3 Single Particle Detectors}

A second SPD system relied on two Timepix3-based hybrid detectors~\cite{Poikela2014} arranged in a stacked geometry, with one 40 mm behind the other. 
The rear was offset laterally by 4 mm, closer to the beam axis.
Both detectors were placed before the CsI SPD.
Timepix3 are CMOS based hybrid pixel detectors with an active area of $14\times14~\textrm{mm}^2$ ($256\times256$ pixels, each \SI{55}{\micro\metre}\,$\times$\,\SI{55}{\micro\metre}).
Timepix3 assemblies with silicon sensors were chosen with the front detector having a \SI{300}{\micro\metre} thick sensor and the rear \SI{150}{\micro\metre} thick.
Any minimum ionising particles (MIPs) passing through the silicon deposit $\sim2900$ eV per micron, registering a trail of charge across the pixels. 
A MIP is considered to be a charged particle that has kinetic energy twice its rest mass, i.e. any positron particles over 1 MeV.
As charged particles pass through the detector, each pixel records the \textit{time-over-threshold} at a rate of up to 640 Mhz.
If fully calibrated the temporal resolution can be as low as $\sim1$ ns, but in our case we integrate the signal over one second during the shot period.
A significant advantage of these detectors is that all charged particles of interest will register a signal, where as due to the silicon nature of the detector, they are relatively insensitive to high energy photons, i.e. they will be insensitive to background $\gamma$ photon noise.
For example, the probability of $\gamma$-ray absorption in \SI{150}{\micro\metre} silicon is $\sim8\times10^{-4}-2\times10^{-3}$~\cite{Hubbell1969}.
As an additional step to confirm that we are witnessing primarily charged particles rather than photons, the stacked Timepix3 setup was simulated in Geant4 with either an incoming particle beam, or an incoming photon beam.
In the experiment calibration shots (discussed further below), an approximately equal number of strikes was measured on the front and rear detectors, but with 0.4 times the total energy absorbed in the rear (the sensor of half the thickness).
This aligns with the Geant4 simulations for charged particle strikes, as opposed to photons which estimated approximately 2.5 times the energy should be deposited in the rear detector due to secondaries created from photon absorption in the first detector.

\begin{figure}
    \centering
    \includegraphics[width=0.95\textwidth]{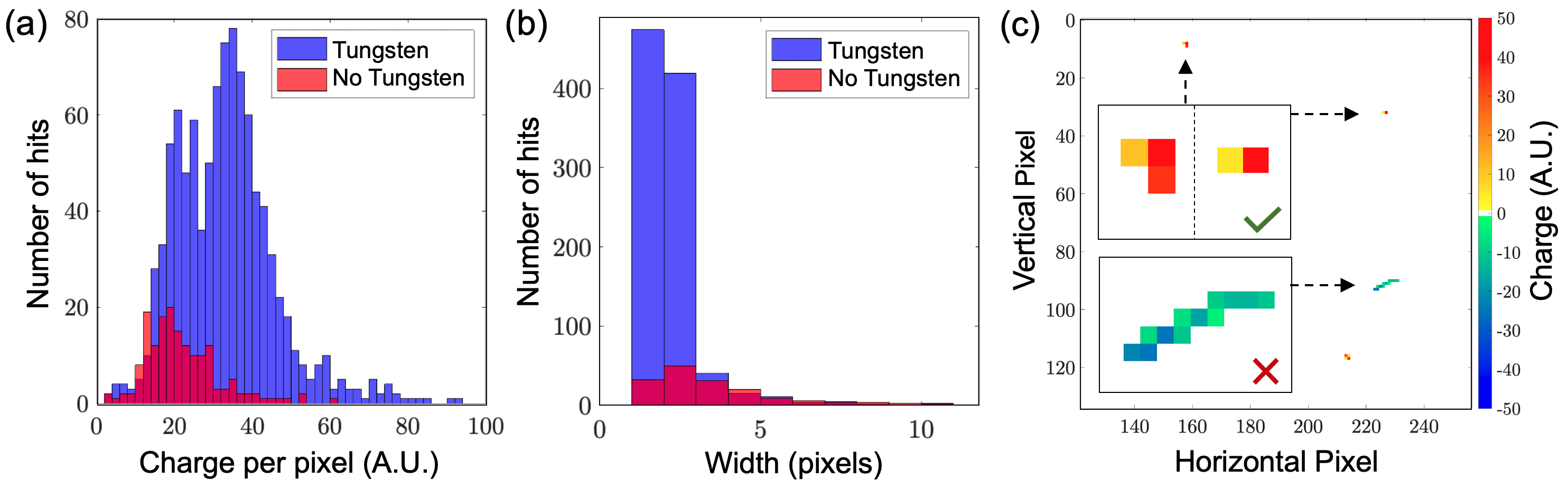}
    \caption{TimePix3 training data. Average charge per pixel (a) and cluster width (b) distributions for shots with and without an on-axis positron source (Tungsten). The difference in the distributions with and without the positron source form the basis of a positron event specification. 
    (c) Examples of the positron specification being implemented and rejecting non-conforming hits (highlighted with a negative colour value).
    }
    \label{fig:TimePixSigs}
\end{figure}

We deem each contiguously connected group of pixel charge registered on the detectors (a ``cluster") as a single charged particle or photon hit.
In order to reduce unwanted noise, we can filter these clusters by their characteristic traits, to select only those that are consistent with positrons that have been transported through the AMS (as opposed to stray background particles).
We do so by training an event identification criteria on a set of calibration data with an artificially created positron signal at the collision point (in a similar manner to the PTFE calibration of the CsI SPDs).
For this, a 1 mm tungsten cube is placed on the $\gamma$ beam axis, within a few cm of the collision point.
The generated particle pairs are similar in energy to those created by our photon-photon collision (100's MeV).
We collated the cluster statistics on 28 shots with the tungsten cube in place, and 28 null shots immediately before this, with the tungsten cube removed.
The latter shots act as a comparable null reference for the background present on the Timepix3.
The shots with the artificial positron source show that the clusters have a higher average charge per pixel, as well as narrower distributions for the width and height, see figure \ref{fig:TimePixSigs} (a) and (b).
These criteria can be applied to all clusters to identify AMS positron events and remove background hits.
To build a complete AMS positron specification, we take the distribution for each metric (e.g. cluster width) individually and subtract the null shot series distribution. 
The limits for that metric are then taken as the 95\% interval of the remaining distribution.
The final AMS positron specification for our setup is given in table~\ref{table:TimePixSigs}.
It was observed on subsequent data and background shots that the cluster filtering was rejecting up to 60\% of the total hits on the Timepix3 detectors.
See figure \ref{fig:TimePixSigs} (c) for an example of this AMS positron identification.  

\begin{table}[!ht]
\centering
\begin{tabular}{ |p{3cm}||p{3.5cm}|p{2.5cm}|p{2cm}|p{2cm}|  }

 Detector & Charge Sum & Num. Pixels & Height & Width \\
 \hhline{|=|=|=|=|=|}
 Front \SI{300}{\micro\metre} & $40 \leq X \leq 145$ & $\leq5$ & $\leq2$ & $\leq3$\\
 Rear \SI{150}{\micro\metre} & $10 \leq X \leq 85$ & $\leq3$ & $\leq2$ & $\leq2$\\
 \hline
\end{tabular}
\caption{Timepix3 on-axis positron specifications. Applying these criteria to each TimePix3 charge cluster allows the identification of positrons travelling through the AMS, and rejects events that are background noise.}
\label{table:TimePixSigs}
\end{table}

The AMS positron shot series was also used to infer the particle trajectory between the two Timepix3 detectors.
With a known stream of incident particles passing through both detectors, it is possible in theory to track the particles as they travel ballistically through the front detector and onto the rear.
To do so, on each shot a 2D map of positron hits was generated for each Timepix3 detector.
A raster of horizontal and vertical positions of one detector relative to the other was then calculated, and a correlation value given to the overlapping region.
Accounting for the fact the rear detector was offset laterally by 4 mm closer to the beam axis, it was found that particles had a range of trajectories of 4$^\circ$-12$^\circ$ between the two detectors. 
This is supported by the Radia tracking simulations for the AMS.
Unfortunately this variation in angular spread (stemming from the variation in particle energy through the AMS) meant it was not possible to track particles from the front Timepix3 through to the rear.
It was possible however to treat the area of the second TimePix3 outside of these trajectories as an additional detection area, without risking double counting positrons passing through the first TimePix3. 
This increased the total detection area to 308 mm$^2$ (a 57\% increase). 
\section{Commissioning} \label{section:commissioning}

\subsection{Characterisation of background sources}

The radiation background was characterised using the CsI SPD and found to be composed of two distinct particle types, high energy positrons transported through the analyser magnet system and low energy diffuse radiation from the $\gamma$-ray beam dump.
The discrimination between these two was made by analysing the yield on different spatial regions on the SPD shown on figure~\ref{fig:simulated-vs-measured-PTFE-positron-distribution-on-SPD}, where AMS positrons can only impact part of the SPD aperture with the remaining area being shielded.
The measured background in the shielded region below the lead brick shadow and right of the AMS cut-off was equivalent to $BG_{diffuse} = 1.3 \pm 0.1  \times 10^{-3}~\mathrm{positrons_{ams}}/\mathrm{pC_r}/\mathrm{crystal}$.
This is assumed to originate from diffuse low energy radiation leaking through the $\gamma$-ray beam dump due to its homogeneous distribution over the detector aperture.
The average background level in the region open for positrons was equivalent to $BG_{open} = 2.5 \pm 0.1  \times 10^{-3}~\mathrm{positrons_{ams}}/\mathrm{pC_r}/\mathrm{crystal}$, almost twice as large compared to the shielded region.
The difference, ($BG_{open} - BG_{diffuse}) = BG_{AMS} = {1.2 \pm 0.1  \times 10^{-3}~\mathrm{positrons_{ams}}/\mathrm{pC_r}/\mathrm{crystal}}$ is therefore assumed to originate from high energy positrons that propagate through the analyser magnet system.
For context, this corresponds to 1 background positron for every 4~pC$_r$ of reference LWFA beam charge integrated over the full 208 crystal large CsI SPD region that is open for AMS positrons.

\begin{figure}
    \centering
    \includegraphics[width=0.95\textwidth]{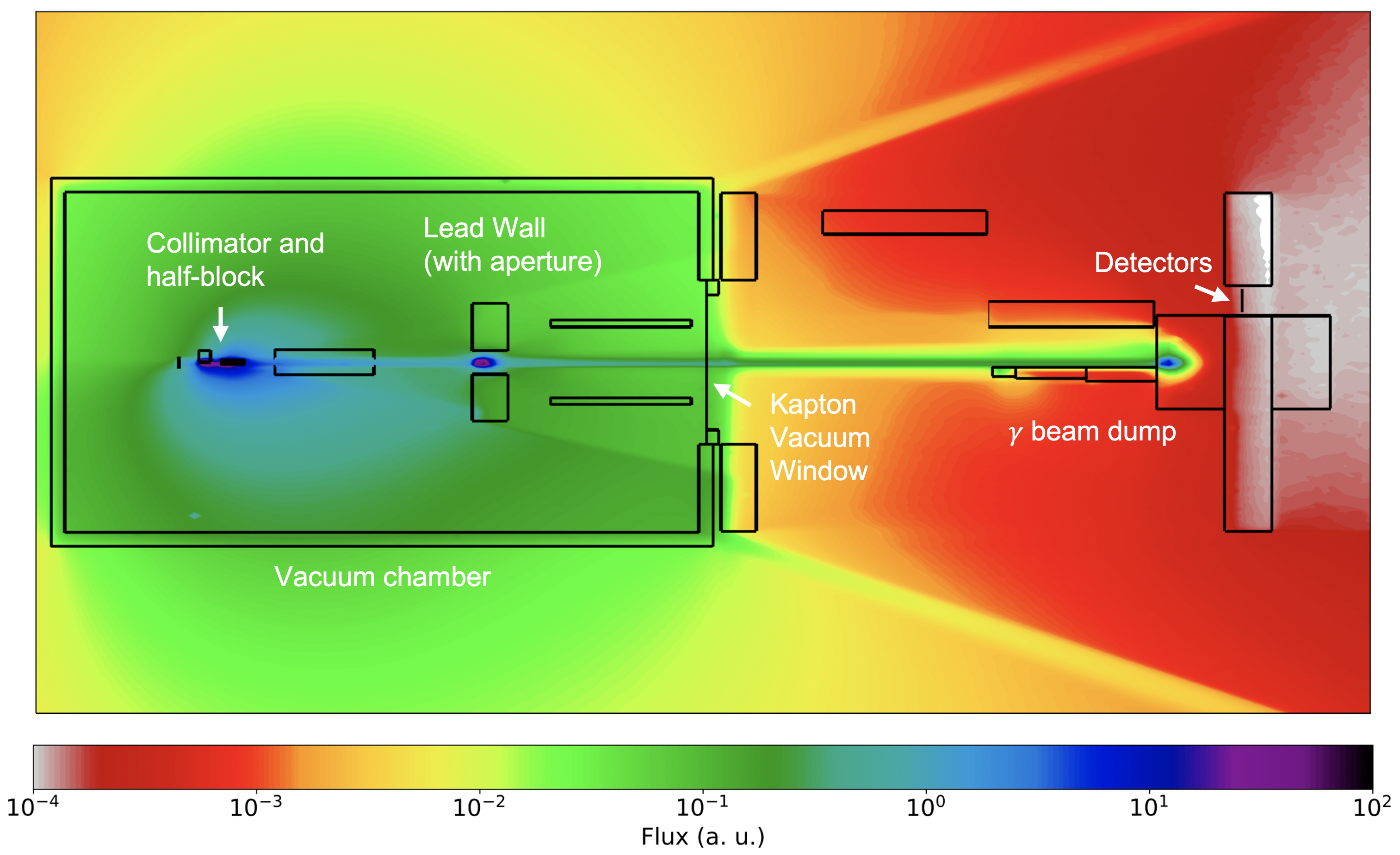}
    \caption{Combined photon and charged particle background noise flux throughout the setup. Simulated in Geant4. Lead walls shield the SPD location from direct sight lines to heavy noise sources such as the $\gamma$ collimator and first lead wall aperture (labelled for clarity).}
    \label{fig:BGLocations}
\end{figure}

Sources of background AMS positrons are Bethe-Heitler pairs produced by the $\gamma$-ray beam in collisions with residual gas inside the vacuum chamber induced by the LWFA target or with apertures close to the beam line, for example the vacuum exit window or lead shielding in between the BW interaction point and the separator magnet shown on figure~\ref{fig:ExpSetup}.
Geant4 simulations were used to understand these noise sources and help design an appropriate shielding configuration. 
Figure~\ref{fig:BGLocations} depicts the combined photon and charged particle flux estimated throughout the final shielded setup.
The main goal of any shielding is to minimise the flux through the SPD location.
The Geant4 simulations identify that the main sources of background positrons which are transported through the AMS are the $\gamma$ beam collimator and the aperture in the lead wall before separator magnet.  
The aperture size was chosen to minimise the background generated by particles (both $\gamma$ rays and electrons from the LWFA) striking the edge of the aperture while maintaining a low background caused by particles striking the pole pieces of the separator magnet.
The primary source of photon background at the detector was identified as the $\gamma$ beam profile and  $\gamma$ spectrum detectors and the beam dump.

\subsection{Gamma dependence of background noise}

To assess the capability of the photon-photon platform to measure the BW process it is important to quantify the background level of the SPDs on data shots.
This measurement was made by taking a series of shots where the $\gamma$ beam was operating and the x-ray foil was at the collision position, but the second laser beam which generates the x-rays was not fired. 
Since all the background sources are generated by the interaction of the $\gamma$ beam with components of the experiment, including the x-ray generation foil, these shots represent a direct measurement of the background that will be present on shots where the two photon sources are colliding. 
We call these type of shots \emph{null shots}.

\begin{figure}
    \centering
    \includegraphics[width=0.45\textwidth]{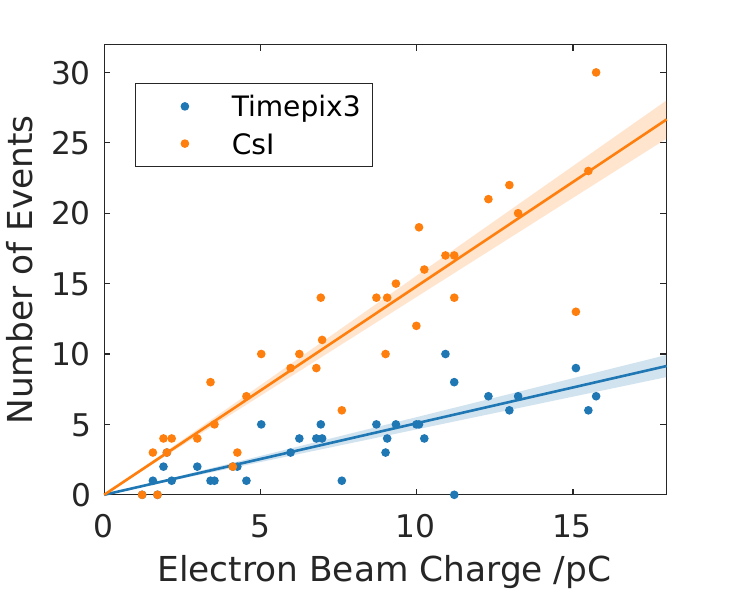}
    \caption{Variation of number of background events measured with electron beam charge on the Timepix3 and CsI detectors. 
    Orange: CsI detector, Blue: Timepix3 detector. 
    Shaded regions represent $1\sigma$ confidence interval on the linear fit.}
    \label{fig:SPDNulls}
\end{figure}

In a series of null shots it was found that the number of background events detected by both SPDs is linearly proportional to the effective charge in the electron beam, as measured on the calibrated $\gamma$ profile diagnostic (see figure~\ref{fig:SPDNulls}).
A linear regression, assuming that the number of background events follows Poisson statistics, with the expected number of background events $\langle N_{\rm bg} \rangle$ varying linearly with the electron charge, $C$,  i.e. 
$\langle N_{\rm bg} \rangle = \sigma_{\rm bg} C$
shows that the signal on the CsI detector corresponds to $ 11.2 \pm 0.6 $ positron events on a shot with the average charge. 
The number of events counted on the Timepix3 detector corresponds to $3.8\pm0.3$ events on an average shot.  
The CsI detector area (5200~mm$^2$) is much larger than the Timepix3 area ($308$~mm$^2$), so we might expect the CsI detector signal to be 17 times higher than the Timepix3 detector.
However the spatial non-uniformity of the background on the CsI detector is expected to reduce this to a ratio of 9. 
The observed ratio is in fact close to 3.

The two detectors have very different sensitivity to the different background sources.  
The CsI detector analysis removes the spatially diffuse background caused by low energy $\gamma$ events generated in the $\gamma$ beam dump. 
This background subtraction is possible on the CsI detector  because part of the detector is shielded from positrons transported by the analyser magnet system.
However, this is not possible for the smaller Timepix3 detector. 
This difference in the background subtraction method on the detectors could be responsible for the difference in the observed rate on null shots. 
However, the thin silicon Timepix3 detector is quite insensitive to $\gamma$ rays (at most $10^{-3}$ absorption), and the track shape detection is able to discount low energy events and events incident at an angle that is inconsistent with positrons transported through the analyser magnet system. 
This suggests that both detectors should be fairly insensitive to the diffuse $\gamma$ background caused by the beam dump. 

\subsection{Commissioning with low yield positron source}

\begin{figure}
    \centering
    \includegraphics[width=0.8\textwidth]{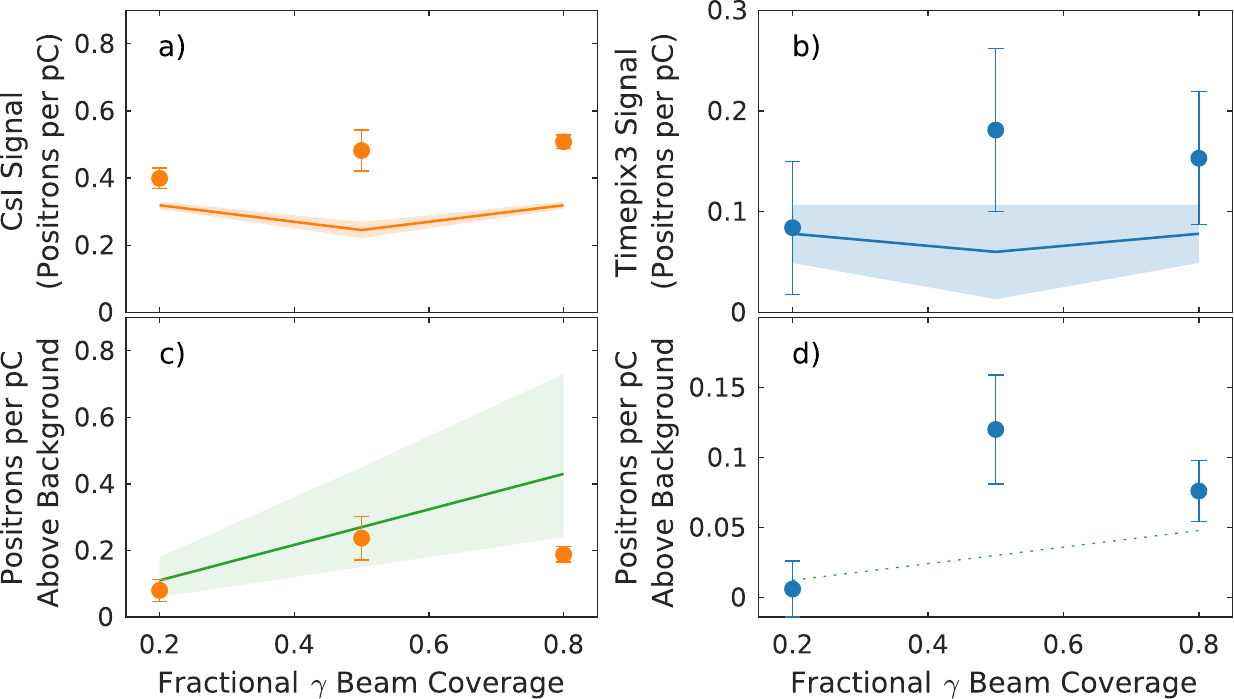}
    \caption{Commissioning of the single particle detectors. 
    Positrons are generated by colliding a fraction of the $\gamma$ beam with a \SI{25}{\micro\metre} thick Kapton target. 
    (a) Positron signal measured (after gamma background subtraction) on the CsI single particle detector. The shaded region is the measured background on a series of shots with the Kapton removed.
    (b) Positron signal measured on the Timepix3 detector.
    The shaded region is the measured background on a series of shots with the Kapton removed.
    (c) Positrons above null shot background measured on the CsI detector. 
    The shaded region is the predicted number of positrons transported to the detector under the same experimental conditions.
    (d) Positrons above null shot background measured on Timepix3 detector. 
    The dashed line is an estimate of predicted number based on scaling the estimated number of positrons on the larger CsI detector.}
    \label{fig:KaptonData}
\end{figure}

To test the detection capability of the SPDs, a series of commissioning shots were performed. 
These involved inserting a \SI{25}{\micro\metre} thick Kapton tape target into the $\gamma$ beam at the photon-photon collision location (while not firing the x-ray drive laser).
The tape target was 1 cm wide, and spooled in the vertical plane, presenting $\approx20$ cm of tape length.
This meant it covered the entire vertical length of the $\gamma$ footprint, but some fraction of the footprint in the horizontal plane.
By translating the target so that it was only partially intercepting the $\gamma$ beam profile we could vary the number of positrons produced and transported to the detectors.
Only a limited number of shots ($\lesssim10$) were taken at each position.  
The $\gamma$ background was subtracted from the CsI data using the method describe in section~\ref{section:detectors}.
The background on both detectors owing to particles generated away from the collision point (e.g. in the lead aperture) being transported to the detectors was characterised by completely removing the Kapton target from the $\gamma$ beam (null shots).  
The measured positron signal is shown in figure  \ref{fig:KaptonData} for both detectors.
After subtraction of the positron background, the CsI detector shows agreement within experimental uncertainty with the predicted number of positron reaching the detector calculated using Geant4, confirming the calibration procedure in this low signal regime. 
The larger area CsI detector was capable of measuring a signal above the positron background when only 20\% of the $\gamma$ beam was intercepted by the Kapton tape, corresponding to  signal level of approximately 0.1 positrons per picocoulomb reaching the detector.
The smaller area Timepix3 detector was able to discern a signal above background when 50\% of the $\gamma$ beam was intercepted, which is also a signal level of approximately 0.1 positrons per picocoulomb reaching the detector.
Even with the limited amount of data taken in these commissioning shots, it is clear that the single particle detectors are capable of measuring small numbers of positrons per picocoulomb.  
However the detection limit of approximately 0.1 positrons per pC above background is very dependent on the number of shots taken to characterise both the signal and the background.
In a data run with many shots we would be able to achieve a significantly higher sensitivity.

\section{Modelling}\label{section:predictions}

The key factor which determines whether a BW detection experiment is successful is the signal-to-noise ratio.
In order to model this parameter and interpret it correctly, accurate predictions involving large scale numerical simulations are required. 
During the design and optimisation phase of this experiment, the Monte-Carlo platform Geant4 was used to perform background noise calculations.
It is desirable to also perform signal calculations within this same framework, enabling faster design and optimisation of the experimental setup.
However, the BW process is not included within any of the standard Geant4 physics packages.
Therefore, a new photon-physics package was developed for this purpose which includes the BW process.
Here a brief overview of the package is provided, however, a more in depth review can be found in reference~\cite{GEANTModulePaper}.

\subsection{Geant4 Breit-Wheeler Module}
Geant4 is a package for simulating the passage of particles through matter.
To include the BW process within Geant4, it must be treated in a similar way.
This has been achieved by handling the two interacting photon sources differently where one is treated as a dynamic beam of photons, and the other is treated as a static field fully defined by a spectral photon density per unit volume per unit solid angle, $n(\omega, \theta, \phi)$. 
Individual photons from the dynamic source are sampled and tracked through the static field.
Calculations are then performed to determine whether the BW process has occurred, and if so, the properties of the emitted electron-positron pair are found.

The interaction between the dynamic particle beam and the static photon field
occurs at a constant average rate making it a Poisson process.
Therefore, the probability that a dynamic particle travels a length, $x$, is given by the following exponential distribution
\begin{equation} \label{eq:exp-dist}
    P(x) =  \lambda_d^{-1} e^{-x / \lambda_d }
\end{equation}
where $\lambda_\mathrm{d}$ is the mean free path of the dynamic particle.
As this is an interaction between massless particles, the mean free path is given by \cite{Weaver1976} 
\begin{equation} \label{eq:MFP}
    \frac{1}{\lambda_d} =
    \displaystyle
    \int_0^{2\pi} \mathrm{d} \phi \int_{0}^\pi \mathrm{d} \theta \int_0^{\infty}  \mathrm{d} \omega \, \sigma\textrm(s) \, n(\omega,  \phi, \theta)\, (1 - \cos{\theta})
\end{equation}
where the BW cross-section is
\begin{equation} 
    \sigma(s) = \frac{\pi r_e^2 (1 - \beta^2)}{2} \Bigg[(3 - \beta^4) \mathrm{log} \frac{1 + \beta}{1 - \beta} - 2\beta(2 - \beta^2) \Bigg],
\end{equation}
$r_e$ is the classical electron radius, $\beta = \sqrt{1 - s^{-1}}$, and $s=(1-\cos\theta_s)E_1 E_2 $ for two photons of energy $E_1$ and $E_2$ interacting  at angle $\theta_s$. 
For each dynamic photon that passes into the static field, $x$ is sampled from equation \ref{eq:exp-dist}.
If this is longer than the static field, it will propagate through unaffected, if it is shorter, then it will propagate $x$ before interacting.

If an interaction occurs, the dynamic photon is removed from the simulation and an electron-positron pair is added.
However, before adding the new particles their momentum must be obtained, which can be defined through each particle's energy $E_\pm$ and polar, $\theta_\pm$ and azimuthal, $\phi_\pm$ scattering angles, where the $\pm$ indicates a positron or electron.
It is easiest to first obtain these properties in the centre-of-mass frame.
Here, both the electron and positron receive an energy of $E_\pm = \sqrt{s} /2$.
The positron polar scattering angle is then obtained by sampling from the differential cross-section for the BW process  
\begin{equation} \label{eq:diffCS}
\frac{d\sigma}{d\Omega} = \frac{r_e^2\beta}{s}\bigg[\frac{1 + 2\beta \mathrm{sin}^2\theta - \beta^4 - \beta^4 \mathrm{sin}^4\theta}{(1 - \beta^2\mathrm{cos}^2\theta)^2}\bigg],
\end{equation}
whereas the positron azimuthal scattering angle is obtained by sampling from the uniform distribution $\mathcal{U}(0, 2 \pi)$.
Due to conservation of momentum, this also defines the scattering angles of the electron, which are given by $\theta_- = \pi - \theta_+$ and $\phi_- = \phi_+ - \pi$.
After applying a Lorentz boost, these properties are obtained in the laboratory frame, allowing the electron-positron pair to be added to the simulation.

This method does not account for either spatial gradients or a temporal evolution of the static photon field.
However, if the full field is constructed by defining multiple sub fields, each with a different $n(\omega, \theta, \phi)$, then spatial gradients can be accounted for.
It is challenging to account for a temporal evolution due to the static nature of the Geant4 computational domain.
Therefore, this module will only provide accurate predictions if the time for the dynamic photons to traverse the static field is much shorter than the characteristic timescale of the static field.
Due to the large difference in the duration of the sources, this is a good approximation for this experiment, where the x-ray source is designated as the static field and the $\gamma$-ray beam is treated dynamically.

\subsection{Predictions}

By developing a BW module for Geant4, start-to-end simulations of this experiment can be performed within a single framework, allowing for efficient calculations of the signal-to-noise ratio.
Such calculations have been carried out using the electron beam and x-ray field measured during the experiment and discussed in sections \ref{section:gammas} and \ref{section:xrays} respectively.
The results of these calculations are summarised in table \ref{table:signal-to-noise}, which show the predicted number of BW signal positrons, background leptons, and background photons which pass through the Timepix3 detectors.
The background has been separated into leptons and photons as the former produces the same signature as the BW signal positrons on the Timepix3 detectors whereas the latter produces a different signature.
Therefore, photon hits can be easily removed as a noise source, whereas lepton hits cannot.
Note that the quoted numbers refer to the number of particles through the TimePix3 detectors, not the larger CsI SPD.
For estimates of the number of particles through the CsI these numbers need to be scaled by the increased area, however, the BW signal is non-uniform over this increased area.
From the Radia simulations described in section \ref{section:transport}, it is estimated that the CsI would cover approximately 9 times more BW signal than the TimePix3.

\begin{table}[b]
\centering
\caption{Predicted number of background particles and signal Breit-Wheeler pairs reaching the TimePix3 detectors, per pC of primary electrons in the LWFA beam. The ``experiment'' row shows predictions using the experimental measurements whereas the ``optimum'' row shows predictions using the best obtained Gemini parameters.}
\begin{tabular}{p{2.8cm} p{3.8cm} p{3.8cm} p{3.8cm}}
\hline \hline
  & BW pairs / pC & BG leptons / pC  & BG photons / pC  \\
 \hline 
Experiment  & $(1.87 \pm 0.69) \times 10^{-6}$ & $(1.97 \pm 0.25) \times 10^{-2}$ & $1.37 \pm 0.06$\\
Optimal  & $(1.37 \pm 0.05) \times 10^{-3}$ & $(10.36 \pm 1.44) \times 10^{-2}$ & $4.06 \pm 0.25$ \\
\hline \hline
\end{tabular}
\label{table:signal-to-noise}
\end{table}

\begin{figure}[!t]
    \centering
    \includegraphics[width=0.95\textwidth]{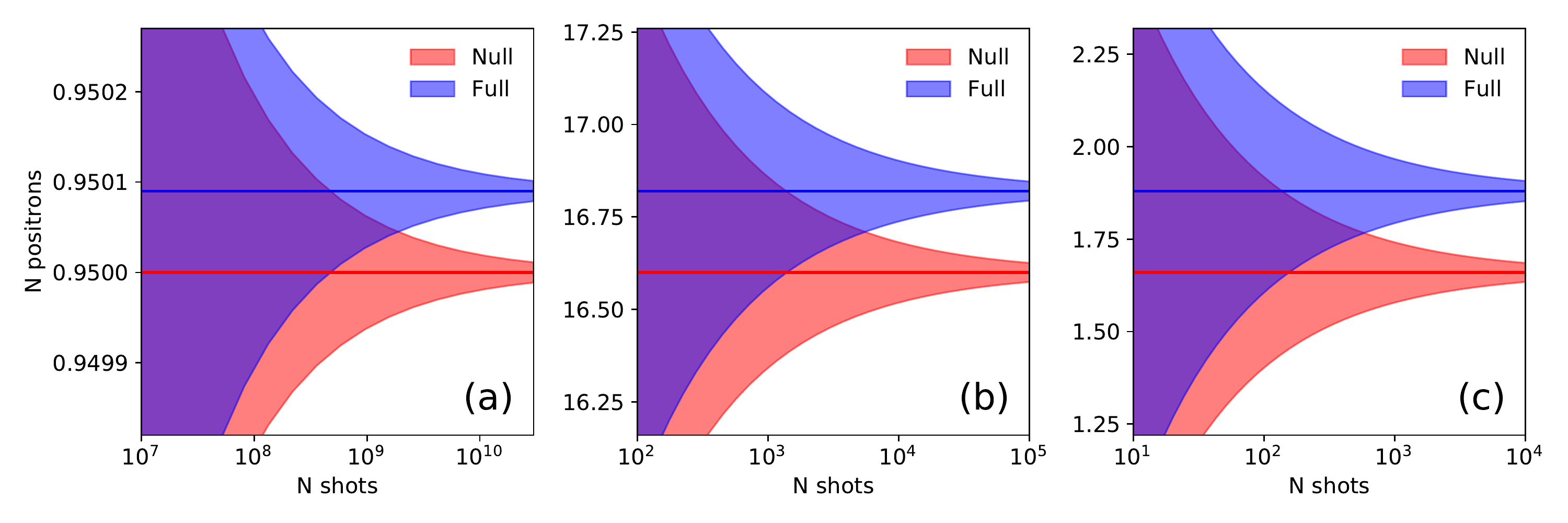}
      \caption{Mean number of positrons and $2\sigma$ error against number of shots, for both full and null shots. (a) Predictions based on measurements made during the experiment. (b) Predictions based on optimum photon sources and the existing shielding configuration. (c) Predictions based on optimum photon sources, but with an extended setup that reduces the background noise by a factor of ten. }
    \label{fig:n-shots}
\end{figure}

To make an observation of the BW process, a statistically significant difference between the mean number of detected positrons on full shots (both laser drivers with $\gamma$ and x-rays colliding) and null shots (no collisions viable as no x-rays present) is required.
Using the values in table \ref{table:signal-to-noise}, a coarse estimate of the number of shots needed to observe this difference can be made.
Assuming an electron beam charge of $50\, \mathrm{pC}$ (corresponding to $\sim 3 \times 10^8$ primary particles), the mean number of events detected on full and null shots are $\mu_\mathrm{F} = 0.95 + (0.9 \times 10^{-4})$ and $\mu_\mathrm{N} = 0.95$ respectively.
The number of events detected on a shot is a Poisson variable, so after $N$ shots the error on $\mu_{\mathrm{N/F}}$ is $\sigma_\mathrm{N/F} = \sqrt{\mu_\mathrm{N/F} / N}$.
If a $2\sigma$ significance observation is desired, then it is required that $\mu_\mathrm{F} - 2 \sigma_\mathrm{F} > \mu_\mathrm{N} + 2 \sigma_\mathrm{N}$.
Figure \ref{fig:n-shots} (a) shows this condition occurs after $\sim 10^{9}$ shots.
This is far more shots than can be achieved over the duration of a Gemini experiment.

\begin{figure*}[t!]
    \centering
        \includegraphics[width=0.5\textwidth]{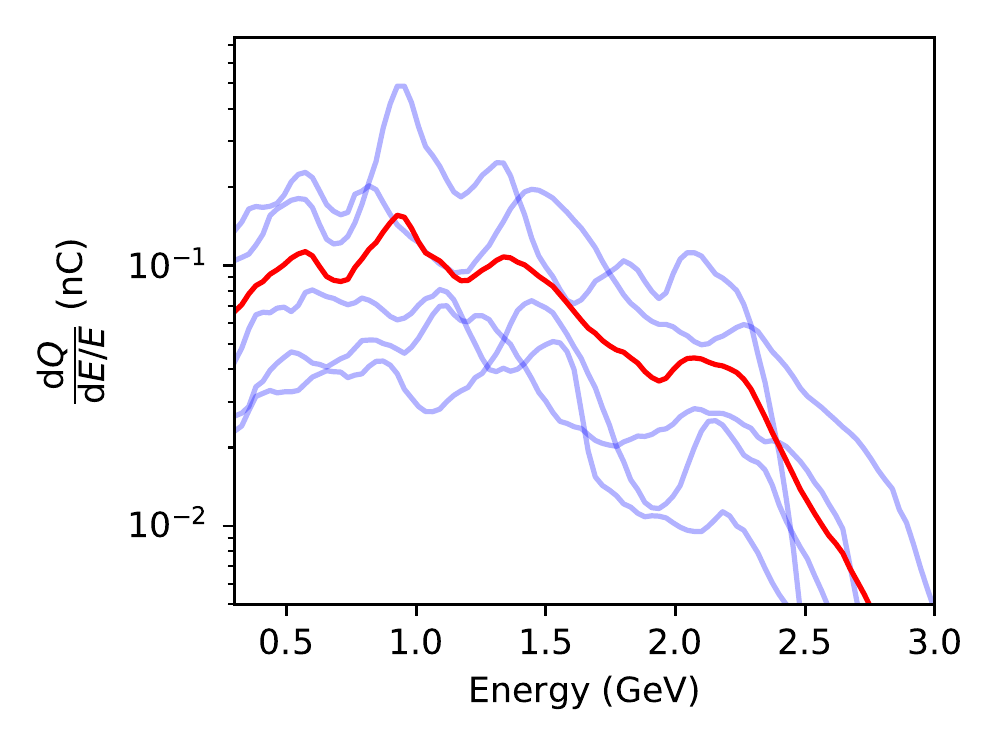}
        \caption{High energy electron beams obtained during a previous Gemini experiment. Taken from ref \cite{PoderThesis}}
        \label{fig:opt-electron-beam}
\end{figure*}

The signal-to-noise ratio obtained using the measured experimental parameters is far lower than what had been predicted from coarse calculations carried out prior to the experiment.
This is due to both the electron beam and x-ray field behaving sub-optimally during the experiment.
For example, figure \ref{fig:opt-electron-beam} shows electron beam spectra obtained during a previous Gemini experiment (see ref \cite{PoderThesis}).
These spectra extend to far higher energies than figure \ref{fig:ElectronPerformance}, with higher beam charges, in excess of $100\, \mathrm{pC}$, also obtained.
On top of this, the energy in the laser pulse used to drive the x-ray field was also lower than expected.
In previous experiments  $15\, \mathrm{J}$ on target has been achieved, a factor of $\sim 1.5$ times higher than this experiment.
Including these factors leads to a significant increase in the BW yield as demonstrated in table \ref{table:signal-to-noise}.
Using the same approach as before, the number of shots required to observe the BW process with optimum conditions can be estimated.
Assuming a beam charge of $160\, \mathrm{pC}$ (corresponding to $10^9$ primary electrons), figure \ref{fig:n-shots} (b) shows  $\sim 5000$ shots are required to make a $2\sigma$ observation with optimum experimental parameters.

\section{Conclusions and Future Directions}

A laser-plasma platform for investigating photon-photon physics has been presented.
A high brilliance $\gamma$ source with associated diagnostics for yield and spectrum was commissioned.
It was found that $(7\pm3)\times 10^{7}$ $\gamma$ photons/shot can be produced with energies over 100 MeV. 
In tandem, a high repetition rate tape-based keV x-ray source was also commissioned.
Approximately $(1.4 \pm 0.5) \times 10^{12}$ x-ray photons/mm$^{3}$ in the energy range 1.3-1.5 keV were produced in the proximity of the photon collision interaction region.
A magnetic transport system was implemented to successfully relay an estimated 14\% of any pair particles produced in the interaction region to two SPD systems (38\% of particles over 200 MeV).
A CsI SPD was tested and was able to detect a signal level of $0.08 \pm 0.03~ \mathrm{positrons_{ams}}/\mathrm{pC_r}$ over a background level equivalent to $0.32 \pm 0.01~ \mathrm{positrons_{ams}}/\mathrm{pC_r}$
This agreed, within uncertainty, with the simulated  signal level.
Additionally a Timepix3 SPD system was also used, recording a signal level of $(0.08 \pm 0.07)~\mathrm{positrons_{ams}}/\mathrm{pC_r}$ over a background of $(0.07 \pm 0.04)~\mathrm{positrons_{ams}}/\mathrm{pC_r}$.

These detection limits where achieved over only $\lesssim10$ shots with the sensitivity expected to increase with the number of shots. 

We estimated that with the achieved performance of the platform, over $10^9$ shots would be needed to observe the linear BW process with $2\sigma$ significance.
However, it is clear that the electron beam performance is crucial to the capabilities of the setup.
Given optimal performance of the laser facility in question, electron beam energies of over 1.5 GeV are possible. 
Simulations indicate that the increased BW signal level associated with this increase in electron energy reduces the number of required shots to approximately 5000.
With the repetition rate of the laser system used, this is achievable in  under 28 hours of operation, which would be an achievable feat.
Note it has been shown recently that the stability and performance of LWFA electrons beams can be improved using machine learning techniques such as Bayesian optimisation~\cite{Shalloo2020, Jalas2021, Hatfield2021}.

In addition, we have identified possible improvements to the experiment setup to significantly increase the efficiency of the platform.
Firstly, the Timepix3 SPD setup in this instance was limited to covering approximately 3 cm$^2$ of the signal area (just over one sensor).
Increasing this detector area with the implementation of more sensors will reduce the number of required shots as more of the pair signal is collected.
For example doubling the number of sensors will halve the amount of data required.
It is estimated that five sensors could cover the majority of the pair signal at the detector plane and reduce the number of required shots for $2\sigma$ significance to $\sim1000$; less than 6 hours of data taking.
With more investment and development it should also be possible to increase the efficiency of the SPDs further.
This might include (but is not limited to) a fully functioning telescope system \cite{Akiba2013} for the Timepix3 detectors to enable particle tracking, or tilting the chips to increase the capability of filtering by angular trajectory \cite{Biagetti2009}.
Fully calibrated Timepix3 sensors (with close to nanosecond time resolution~\cite{Pitters2019}) could also employ time-of-flight distinction between signal and noise sources.
An upgraded Timepix4 sensor will also soon be available which not only has a larger detection area ($\sim$7~ cm$^2$ area) but can be interconnected to tile large areas~\cite{TimePix4}.

Secondly, by making the setup more compact and bringing the interaction point closer to the $\gamma$ source, a greater fraction of the $\gamma$ rays can overlap with the regions of highest $x$ ray density, increasing the pair yield. 
Alternatively an active plasma lens~\cite{Lindstrom2018} could be used to focus the electron beam, creating a narrower $\gamma$-ray beam at the collision point, allowing a greater fraction of the $\gamma$ rays to interact with the regions of highest $x$-ray density.

Finally it is clear that reducing the detector background noise can be as important as increasing the number of pairs created or the platform efficiency.
The number of shots required to make a significant measurement scales $\sim$ noise / signal$^2$. 
Undertaking a more detailed design study or shielding construction to reduce the detector noise could have a major impact on the length of time that needs to be dedicated to data taking.
As an example, figure \ref{fig:n-shots} (c) depicts the number of shots required to make a $2\sigma$ significance measurement with the optimum performance presented previously, but with a x10 reduction in background level.
The number of shots required is now only approximately 500.

It should be noted that future opportunities will be provided by increased laser capabilities coming online in the near future at other facilities across the globe~\cite{EPACLaser, ZeusLaser, ApollonLaser, ELILaser, OmegaEPOpalLaser, CoReLSLaser}.
For example, with 1 PW it is possible to produce electron beams of over 8 GeV energy~\cite{Gonsalves2019}.
As described previously the achievable electron energy is critical to the production rate of the pairs.
The planned facilities are also in general closer to a repetition rate of 1 Hz, and will have increased laboratory space.
The magnetic transport system and SPDs described for this platform were designed to fit within the size constraints of the Gemini target area which is less than 8\,m long.
With more available space it would be possible to separate the electron and $\gamma$ beam dumps from the SPDs, and to introduce more shielding, reducing the background level further, which as mentioned above plays a major role in reducing the number of shots required for a measurement.
More space would remove limitations in the spectral selection and focusing capabilities of the magnetic transport system.

These developments of the platform lead to exciting opportunities for investigating other photon-photon physics.
This includes testing the nonlinear BW process by replacing the x-ray bath source with the direct high-intensity laser pulse.
It may also be possible to use the system to measure or reduce the upper bound on the  cross section for photon-photon scattering \cite{Bernard2000, Yamaji2016}.


\section*{Acknowledgements}

We wish to acknowledge the support of the staff at the Central Laser Facility, the help from Robin Pitman and Vernon Gibson, and also to thank the Medipix3 Collaboration for providing the Timepix3 sensors.
This project has received funding from  the  European  Research  Council  (ERC)  under  the European Union’s Horizon 2020 research and innovation programme (grant agreement no 682399)
and STFC (grant No: ST/P002021/1).
JH and AT acknowledge support from the US National Science Foundation grant \#1804463.
GS would like to acknowledge support from EPSRC (grant No: EP/N027175/1, EP/P010059/1).

\section*{References}
\bibliographystyle{iopart-num}
\bibliography{references}

\providecommand{\newblock}{}
\begin{thebibliography}{10}
\expandafter\ifx\csname url\endcsname\relax
  \def\url#1{{\tt #1}}\fi
\expandafter\ifx\csname urlprefix\endcsname\relax\def\urlprefix{URL }\fi
\providecommand{\eprint}[2][]{\url{#2}}

\bibitem{Breit1934}
Breit G and Wheeler J~A 1934 {\em Physical Review\/} {\bf 46}(12) 1087--1091
  ISSN 0031899X

\bibitem{greiner2008quantum}
Greiner W and Reinhardt J 2008 {\em Quantum electrodynamics\/} (Springer
  Science \& Business Media)

\bibitem{Asner2003}
Asner D, Burkhardt H, De~Roeck A, Ellis J, Gronberg J, Heinemeyer S, Schmitt M,
  Schulte D, Velasco M and Zimmermann F 2003 {\em Eur. Phys. J. C\/} {\bf 28}
  27--44 \urlprefix\url{https://doi.org/10.1140/epjc/s2002-01113-3}

\bibitem{Takahashi2019}
Takahashi T 2019 {\em Reviews of Accelerator Science and Technology\/} {\bf 10}
  215--226 \urlprefix\url{https://doi.org/10.1142/S1793626819300111}

\bibitem{Kneip2009}
Kneip S, Nagel S~R, Martins S~F, Mangles S~P~D, Bellei C, Chekhlov O, Clarke
  R~J, Delerue N, Divall E~J, Doucas G, Ertel K, Fiuza F, Fonseca R, Foster P,
  Hawkes S~J, Hooker C~J, Krushelnick K, Mori W~B, Palmer C~A~J, Phuoc K~T,
  Rajeev P~P, Schreiber J, Streeter M~J~V, Urner D, Vieira J, Silva L~O and
  Najmudin Z 2009 {\em Phys. Rev. Lett.\/} {\bf 103}(3) 035002
  \urlprefix\url{https://link.aps.org/doi/10.1103/PhysRevLett.103.035002}

\bibitem{Gonsalves2019}
Gonsalves A~J, Nakamura K, Daniels J, Benedetti C, Pieronek C, de~Raadt T~C~H,
  Steinke S, Bin J~H, Bulanov S~S, van Tilborg J, Geddes C~G~R, Schroeder C~B,
  T\'oth C, Esarey E, Swanson K, Fan-Chiang L, Bagdasarov G, Bobrova N, Gasilov
  V, Korn G, Sasorov P and Leemans W~P 2019 {\em Phys. Rev. Lett.\/} {\bf
  122}(8) 084801
  \urlprefix\url{https://link.aps.org/doi/10.1103/PhysRevLett.122.084801}

\bibitem{Glinec2005}
Glinec Y, Faure J, Dain L~L, Darbon S, Hosokai T, Santos J~J, Lefebvre E,
  Rousseau J~P, Burgy F, Mercier B and Malka V 2005 {\em Phys. Rev. Lett.\/}
  {\bf 94} 025003
  \urlprefix\url{https://link.aps.org/doi/10.1103/PhysRevLett.94.025003}

\bibitem{lemos2018bremsstrahlung}
Lemos N, Albert F, Shaw J, Papp D, Polanek R, King P, Milder A, Marsh K, Pak A,
  Pollock B {\em et~al.\/} 2018 {\em Plasma Physics and Controlled Fusion\/}
  {\bf 60} 054008

\bibitem{dopp2016bremsstrahlung}
D{\"o}pp A, Guillaume E, Thaury C, Lifschitz A, Sylla F, Goddet J~P, Tafzi A,
  Iaquanello G, Lefrou T, Rousseau P {\em et~al.\/} 2016 {\em Nuclear
  Instruments and Methods in Physics Research Section A: Accelerators,
  Spectrometers, Detectors and Associated Equipment\/} {\bf 830} 515--519

\bibitem{underwood2020development}
Underwood C~I~D, Baird C~D, Murphy C, Armstrong C, Thornton C, Finlay O,
  Streeter M~J, Selwood M~P, Brierley N, Cipiccia S {\em et~al.\/} 2020 {\em
  Plasma Physics and Controlled Fusion\/} {\bf 62} 124002

\bibitem{Sarri2014}
Sarri G, Corvan D, Schumaker W, Cole J, Di~Piazza A, Ahmed H, Harvey C, Keitel
  C~H, Krushelnick K, Mangles S and M Z 2014 {\em Physical Review Letters\/}
  {\bf 113} 224801

\bibitem{Yan2017}
Yan W, Fruhling C, Golovin G, Haden D, Luo J, Zhang P, Zhao B, Zhang J, Liu C,
  Chen M {\em et~al.\/} 2017 {\em Nature Photonics\/} {\bf 11} 514--520

\bibitem{Cole2018}
Cole J, Behm K, Gerstmayr E, Blackburn T, Wood J, Baird C, Duff M~J, Harvey C,
  Ilderton A, Joglekar A {\em et~al.\/} 2018 {\em Physical Review X\/} {\bf 8}
  011020

\bibitem{Pike2014}
Pike O~J, MacKenroth F, Hill E~G and Rose S~J 2014 {\em Nature Photonics\/}
  {\bf 8}(6) 434--436 ISSN 17494893
  \urlprefix\url{http://dx.doi.org/10.1038/nphoton.2014.95}

\bibitem{golub2021linear}
Golub A, Villalba-Ch{\'a}vez S, Ruhl H and M{\"u}ller C 2021 {\em Physical
  Review D\/} {\bf 103} 016009

\bibitem{Ruffini2010}
Ruffini R, Vereshchagin G and Xue S~S 2010 {Electron-positron pairs in physics
  and astrophysics: From heavy nuclei to black holes} (\textit{Preprint}
  \eprint{0910.0974})
  \urlprefix\url{https://ui.adsabs.harvard.edu/abs/2010PhR...487....1R/abstract}

\bibitem{Rakavy1967}
Rakavy G and Shaviv G 1967 {\em The Astrophysical Journal\/} {\bf 148} 803 ISSN
  0004-637X
  \urlprefix\url{https://ui.adsabs.harvard.edu/abs/1967ApJ...148..803R/abstract}

\bibitem{Barkat1967}
Barkat Z, Rakavy G and Sack N 1967 {\em Physical Review Letters\/} {\bf 18}
  379--381 ISSN 00319007
  \urlprefix\url{https://ui.adsabs.harvard.edu/abs/1967PhRvL..18..379B/abstract}

\bibitem{Pan2012}
Pan T, Kasen D and Loeb A 2012 {\em Monthly Notices of the Royal Astronomical
  Society\/} {\bf 422} 2701--2711 ISSN 00358711 (\textit{Preprint}
  \eprint{1112.2710})
  \urlprefix\url{https://academic.oup.com/mnras/article/422/3/2701/1049879
  https://academic.oup.com/mnras/article-lookup/doi/10.1111/j.1365-2966.2012.20837.x}

\bibitem{Regos2020}
Regős E, Vink{\'{o}} J and Ziegler B~L 2020 {\em The Astrophysical Journal\/}
  {\bf 894} 94 ISSN 1538-4357 (\textit{Preprint} \eprint{2002.07854})
  \urlprefix\url{https://ui.adsabs.harvard.edu/abs/2020ApJ...894...94R/abstract
  https://iopscience.iop.org/article/10.3847/1538-4357/ab8636}

\bibitem{Bonometto1971}
Bonometto S and Rees M~J 1971 {\em Monthly Notices of the Royal Astronomical
  Society\/} {\bf 152} 21--35 ISSN 0035-8711
  \urlprefix\url{https://ui.adsabs.harvard.edu/abs/1971MNRAS.152...21B/abstract
  https://academic.oup.com/mnras/article-lookup/doi/10.1093/mnras/152.1.21}

\bibitem{Fabian2015}
Fabian A~C, Lohfink A, Kara E, Parker M~L, Vasudevan R and Reynolds C~S 2015
  {\em Monthly Notices of the Royal Astronomical Society\/} {\bf 451}
  4375--4383 ISSN 0035-8711 (\textit{Preprint} \eprint{1505.07603})
  \urlprefix\url{https://ui.adsabs.harvard.edu/abs/2015MNRAS.451.4375F/abstract
  https://academic.oup.com/mnras/article-lookup/doi/10.1093/mnras/stv1218}

\bibitem{Gould1966}
Gould R~J and Schr{\'{e}}der G 1966 {\em Physical Review Letters\/} {\bf 16}
  252--254 ISSN 0031-9007
  \urlprefix\url{https://ui.adsabs.harvard.edu/abs/1966PhRvL..16..252G/abstract
  https://link.aps.org/doi/10.1103/PhysRevLett.16.252}

\bibitem{Horns2012}
Horns D, MacCione L, Meyer M, Mirizzi A, Montanino D and Roncadelli M 2012 {\em
  Physical Review D - Particles, Fields, Gravitation and Cosmology\/} {\bf 86}
  075024 ISSN 15507998 (\textit{Preprint} \eprint{1207.0776})
  \urlprefix\url{http://arxiv.org/abs/1207.0776
  http://dx.doi.org/10.1103/PhysRevD.86.075024
  https://link.aps.org/doi/10.1103/PhysRevD.86.075024}

\bibitem{Galanti2020}
Galanti G, Tavecchio F and Landoni M 2020 {\em Monthly Notices of the Royal
  Astronomical Society\/} {\bf 491} 5268--5276 ISSN 0035-8711
  (\textit{Preprint} \eprint{1911.09056})
  \urlprefix\url{https://ui.adsabs.harvard.edu/abs/2020MNRAS.491.5268G/abstract
  https://academic.oup.com/mnras/article/491/4/5268/5685981}

\bibitem{Biteau2015}
Biteau J and Williams D~A 2015 {\em The Astrophysical Journal\/} {\bf 812} 60
  ISSN 1538-4357 (\textit{Preprint} \eprint{1502.04166})
  \urlprefix\url{http://tevcat.uchicago.edu/
  https://iopscience.iop.org/article/10.1088/0004-637X/812/1/60}

\bibitem{Piran2004}
Piran T 2005 {\em Reviews of Modern Physics\/} {\bf 76} 1143--1210 ISSN
  0034-6861 (\textit{Preprint} \eprint{0405503})
  \urlprefix\url{https://ui.adsabs.harvard.edu/abs/2004RvMP...76.1143P/abstract
  https://link.aps.org/doi/10.1103/RevModPhys.76.1143}

\bibitem{Blandford1977}
Blandford R~D and Znajek R~L 1977 {\em Mon. Not. R. Astron. Soc.\/} {\bf 179}
  433--456

\bibitem{Bula1996}
Bula C, McDonald K~T, Prebys E~J, Bamber C, Boege S, Kotseroglou T, Melissinos
  A~C, Meyerhofer D~D, Ragg W, Burke D~L, Field R~C, Horton-Smith G, Odian A~C,
  Spencer J~E, Walz D {\em et~al.\/} 1996 {\em Phys. Rev. Lett.\/} {\bf 76}
  3116

\bibitem{Burke1997}
Burke D~L, Field R~C, Horton-Smith G, Spencer J~E, Walz D, Berridge S~C, Bugg
  W~M, Shmakov K, Weidemann A~W, Bula C, Donald K~T~M, Prebys E~J, Bamber C,
  Boege S~J, Koffas T, Kotseroglou T, Melissinos A~C, Meyerhofer D~D, Reis D~A
  and Ragg W 1997 {\em Physical Review Letters\/} {\bf 79}(9) 1626--1629 ISSN
  10797114

\bibitem{Bamber2004}
Bamber C, Boege S~J, Koffas T, Kotseroglou T, Melissinos A~C, Meyerhofer D~D,
  Reis D~A, Ragg W, Bula C, McDonald K~T, Prebys E~J, Burke D~L, Field R~C,
  Horton-Smith G, Spencer J~E, Walz D, Berridge S~C, Bugg W~M, Shmakov K and
  Weidemann A~W 1999 {\em Phys. Rev. D\/} {\bf 60}(9) 092004
  \urlprefix\url{https://link.aps.org/doi/10.1103/PhysRevD.60.092004}

\bibitem{Ellis1992}
 1992 {\em Nuclear Physics B\/} {\bf 373} 399--437 ISSN 0550-3213
  \urlprefix\url{https://www.sciencedirect.com/science/article/pii/055032139290438H}

\bibitem{Svensson1990}
Svensson R and Zdziarski A 1990 {\em The Astrophysical Journal\/} {\bf 349} 415
  ISSN 0004-637X

\bibitem{Schumacher1975}
Schumacher M, Borchert I, Smend F and Rullhusen P 1975 {\em Physics Letters
  B\/} {\bf 59} 134--136 ISSN 0370-2693
  \urlprefix\url{https://www.sciencedirect.com/science/article/pii/0370269375906851}

\bibitem{ATLAS2017}
Collaboration A 2017 {\em Nature Phys\/} {\bf 13} 852–858
  \urlprefix\url{https://doi.org/10.1038/nphys4208}

\bibitem{Miller2004NIF}
Miller G~H, Moses E~I and Wuest C~R 2004 {\em Optical Engineering\/} {\bf 43}
  2841--2853

\bibitem{Ribeyre2016}
Ribeyre X, d'Humi{\`e}res E, Jansen O, Jequier S, Tikhonchuk V and Lobet M 2016
  {\em Physical Review E\/} {\bf 93} 013201

\bibitem{Drebot2017PRAB}
Drebot I, Micieli D, Milotti E, Petrillo V, Tassi E and Serafini L 2017 {\em
  Physical Review Accelerators and Beams\/} {\bf 20}(4) 1--6 ISSN 24699888

\bibitem{Keitel_Gemini2010}
Keitel C, DiPiazza A, Paulus G, Stoehlker T, Clark E, Mangles S, Najmudin Z,
  Krushelnick K, Schreiber J, Borghesi M, Dromey B, Riley D, Sarri G and Zepf M
  2010 {\em RAL Proposal Reference No: HPL 112008 (2010)\/}
  \urlprefix\url{https://arxiv.org/abs/2103.06059}

\bibitem{corvan2016optical}
Corvan D, Dzelzainis T, Hyland C, Nersisyan G, Yeung M, Zepf M and Sarri G 2016
  {\em Optics express\/} {\bf 24} 3127--3136

\bibitem{DataRepo}
 {\em Data Repository for the experiment\/}
  \urlprefix\url{https://doi.org/10.5281/zenodo.5027591}

\bibitem{TajimaDawson1979}
Tajima T and Dawson J~M 1979 {\em Phys. Rev. Lett.\/} {\bf 43}(4) 267--270
  \urlprefix\url{https://link.aps.org/doi/10.1103/PhysRevLett.43.267}

\bibitem{Behm2018}
Behm K, Cole J, Joglekar A, Gerstmayr E, Wood J, Baird C, Blackburn T, Duff M,
  Harvey C, Ilderton A {\em et~al.\/} 2018 {\em Review of Scientific
  Instruments\/} {\bf 89} 113303

\bibitem{Kato1984}
Kato Y, Mima K, Miyanaga N, Arinaga S, Kitagawa Y, Nakatsuka M and Yamanaka C
  1984 {\em Phys. Rev. Lett.\/} {\bf 53}(11) 1057--1060
  \urlprefix\url{https://link.aps.org/doi/10.1103/PhysRevLett.53.1057}

\bibitem{SciTechTape}
Astbury S, Spindloe C, Tolley M, Harman L, Sykes P, Sarasola R, Rodgers K and
  Robins W 2019 {\em CLF Annual Report\/}
  \urlprefix\url{https://www.clf.stfc.ac.uk/Gallery/33\%20-\%20Astbury.pdf}

\bibitem{PhillionHailey1986}
Phillion D~W and Hailey C~J 1986 {\em Phys. Rev. A\/} {\bf 34}(6) 4886--4896
  \urlprefix\url{https://link.aps.org/doi/10.1103/PhysRevA.34.4886}

\bibitem{Kettle2015}
Kettle B, Dzelzainis T, White S, Li L, Rigby A, Spindloe C, Notley M, Heathcote
  R, Lewis C~L~S and Riley D 2015 {\em Journal of Physics B: Atomic, Molecular
  and Optical Physics\/} {\bf 48} 224002
  \urlprefix\url{https://doi.org/10.1088/0953-4075/48/22/224002}

\bibitem{elleaume1997computing}
Elleaume P, Chubar O and Chavanne J 1997 Computing 3d magnetic fields from
  insertion devices {\em Proceedings of the 1997 Particle Accelerator
  Conference (Cat. No. 97CH36167)\/} vol~3 (IEEE) pp 3509--3511

\bibitem{chubar1998three}
Chubar O, Elleaume P and Chavanne J 1998 {\em Journal of synchrotron
  radiation\/} {\bf 5} 481--484

\bibitem{Poikela2014}
Poikela T, Plosila J, Westerlund T, Campbell M, Gaspari M~D, Llopart X, Gromov
  V, Kluit R, van Beuzekom M, Zappon F, Zivkovic V, Brezina C, Desch K, Fu Y
  and Kruth A 2014 {\em Journal of Instrumentation\/} {\bf 9} C05013--C05013
  \urlprefix\url{https://doi.org/10.1088/1748-0221/9/05/c05013}

\bibitem{Hubbell1969}
Hubbell J 1969 {\em National Bureau of Standards Report NSRDS-NBS29, Washington
  DC\/}

\bibitem{GEANTModulePaper}
Watt R 2021 {\em In Preparation\/}

\bibitem{Weaver1976}
Weaver T 1976 {\em Astrophysical Journal Supplement Series\/} {\bf 32} 233--282

\bibitem{PoderThesis}
Poder K 2016 {\em Characterisation of self-guided laser wakefield accelerators
  to multi-GeV energies\/} Ph.D. thesis
  \urlprefix\url{https://doi.org/10.25560/56216}

\bibitem{Shalloo2020}
Shalloo R~J, Dann S~J~D, Gruse J~N, Underwood C~I~D, Antoine A~F, Arran C,
  Backhouse M, Baird C~D, Balcazar M~D, Bourgeois N, Cardarelli J~A, Hatfield
  P, Kang J, Krushelnick K, Mangles S~P~D, Murphy C~D, Lu N, Osterhoff J,
  P{\~o}der K, Rajeev P~P, Ridgers C~P, Rozario S, Selwood M~P, Shahani A~J,
  Symes D~R, Thomas A~G~R, Thornton C, Najmudin Z and Streeter M~J~V 2020 {\em
  Nature Communications\/} {\bf 11} 6355 ISSN 2041-1723
  \urlprefix\url{https://doi.org/10.1038/s41467-020-20245-6}

\bibitem{Jalas2021}
Jalas S, Kirchen M, Messner P, Winkler P, H\"ubner L, Dirkwinkel J, Schnepp M,
  Lehe R and Maier A~R 2021 {\em Phys. Rev. Lett.\/} {\bf 126}(10) 104801
  \urlprefix\url{https://link.aps.org/doi/10.1103/PhysRevLett.126.104801}

\bibitem{Hatfield2021}
Hatfield P~W, Gaffney J~A, Anderson G~J, Ali S, Antonelli L,
  Ba{\c{s}}e{\u{g}}mez~du Pree S, Citrin J, Fajardo M, Knapp P, Kettle B,
  Kustowski B, MacDonald M~J, Mariscal D, Martin M~E, Nagayama T, Palmer C~A~J,
  Peterson J~L, Rose S, Ruby J~J, Shneider C, Streeter M~J~V, Trickey W and
  Williams B 2021 {\em Nature\/} {\bf 593} 351--361 ISSN 1476-4687
  \urlprefix\url{https://doi.org/10.1038/s41586-021-03382-w}

\bibitem{Akiba2013}
Akiba K, Ronning P, {van Beuzekom} M, {van Beveren} V, Borghi S, Boterenbrood
  H, Buytaert J, Collins P, {Dosil Suárez} A, Dumps R, Eklund L, Esperante D,
  Gallas A, Gordon H, {van der Heijden} B, Hombach C, Hynds D, John M, Leflat
  A, Li Y, Longstaff I, Morton A, Nakatsuka N, Nomerotski A, Parkes C, {Perez
  Trigo} E, Plackett R, Reid M, {Rodriguez Perez} P, Schindler H, Szumlak T,
  Tsopelas P, {Vázquez Sierra} C, Velthuis J and Wysokiński M 2013 {\em
  Nuclear Instruments and Methods in Physics Research Section A: Accelerators,
  Spectrometers, Detectors and Associated Equipment\/} {\bf 723} 47--54 ISSN
  0168-9002
  \urlprefix\url{https://www.sciencedirect.com/science/article/pii/S0168900213004816}

\bibitem{Biagetti2009}
Biagetti D, Meroli S, Marras A, Passeri D, Placidi P and Servoli L 2009 Tilted
  cmos active pixel sensors for particle track reconstruction {\em 2009 IEEE
  Nuclear Science Symposium Conference Record (NSS/MIC)\/} pp 1678--1681

\bibitem{Pitters2019}
Pitters F, Tehrani N~A, Dannheim D, Fiergolski A, Hynds D, Klempt W, Llopart X,
  Munker M, Nürnberg A, Spannagel S and Williams M 2019 {\em Journal of
  Instrumentation\/} {\bf 14} P05022--P05022
  \urlprefix\url{https://doi.org/10.1088/1748-0221/14/05/p05022}

\bibitem{TimePix4}
  \urlprefix\url{https://medipix.web.cern.ch/medipix4}

\bibitem{Lindstrom2018}
Lindstr\o{}m C~A, Adli E, Boyle G, Corsini R, Dyson A~E, Farabolini W, Hooker
  S~M, Meisel M, Osterhoff J, R\"ockemann J~H, Schaper L and Sjobak K~N 2018
  {\em Phys. Rev. Lett.\/} {\bf 121}(19) 194801
  \urlprefix\url{https://link.aps.org/doi/10.1103/PhysRevLett.121.194801}

\bibitem{EPACLaser}
  \urlprefix\url{https://www.clf.stfc.ac.uk/Pages/EPAC-introduction-page.aspx}

\bibitem{ZeusLaser}
  \urlprefix\url{https://zeus.engin.umich.edu/}

\bibitem{ApollonLaser}
Zou J, Le~Blanc C, Papadopoulos D, Ch{\'e}riaux G, Georges P, Mennerat G, Druon
  F, Lecherbourg L, Pellegrina A, Ramirez P {\em et~al.\/} 2015 {\em High Power
  Laser Science and Engineering\/} {\bf 3}

\bibitem{ELILaser}
Gales S, Tanaka K~A, Balabanski D~L, Negoita F, Stutman D, Tesileanu O, Ur C~A,
  Ursescu D, Andrei I, Ataman S, Cernaianu M~O, D'Alessi L, Dancus I,
  Diaconescu B, Djourelov N, Filipescu D, Ghenuche P, Ghita D~G, Matei C, Seto
  K, Zeng M and Zamfir N~V 2018 {\em Reports on Progress in Physics\/} {\bf 81}
  094301 \urlprefix\url{https://doi.org/10.1088/1361-6633/aacfe8}

\bibitem{OmegaEPOpalLaser}
Bromage J, Bahk S~W, Begishev I, Dorrer C, Guardalben M, Hoffman B, Oliver J,
  Roides R, Schiesser E, Shoup~III M {\em et~al.\/} 2019 {\em High Power Laser
  Science and Engineering\/} {\bf 7}

\bibitem{CoReLSLaser}
Yoon J~W, Kim Y~G, Choi I~W, Sung J~H, Lee H~W, Lee S~K and Nam C~H 2021 {\em
  Optica\/} {\bf 8} 630--635
  \urlprefix\url{http://www.osapublishing.org/optica/abstract.cfm?URI=optica-8-5-630}

\bibitem{Bernard2000}
Bernard D, Moulin F, Amiranoff F, Braun A, Chambaret J, Darpentigny G, Grillon
  G, Ranc S and Perrone F 2000 {\em The European Physical Journal D-Atomic,
  Molecular, Optical and Plasma Physics\/} {\bf 10} 141--145

\bibitem{Yamaji2016}
Yamaji T, Inada T, Yamazaki T, Namba T, Asai S, Kobayashi T, Tamasaku K, Tanaka
  Y, Inubushi Y, Sawada K {\em et~al.\/} 2016 {\em Physics Letters B\/} {\bf
  763} 454--457

\end{thebibliography}

\end{document}